\documentclass[10pt]{IEEEtran}

\ifCLASSINFOpdf
\else
\fi

\makeatletter
\def\endthebibliography{%
	\def\@noitemerr{\@latex@warning{Empty `thebibliography' environment}}%
	\endlist
}
\makeatother

\usepackage{multirow}
\usepackage{colortbl}
\usepackage{multicol}
\usepackage{booktabs}
\usepackage{graphicx}
\usepackage{cite}
\usepackage{bbm}
\usepackage{amsfonts}
\usepackage{amsmath}
\usepackage{algorithm}
\usepackage{algorithmic}

\usepackage{amsthm}
\usepackage{MnSymbol}%
\usepackage{wasysym}%
\usepackage{verbatim}
\usepackage{clrscode}

\usepackage{subfigure}
\usepackage{float}
\usepackage{makecell}
\usepackage[labelfont=bf]{caption}
\usepackage{color}
\usepackage{rotating}

\IEEEoverridecommandlockouts

\begin{document}
%
\title{Can We Trust Your Explanations? Sanity Checks for Interpreters in Android Malware Analysis}

\author{Ming Fan\IEEEauthorrefmark{1}\IEEEauthorrefmark{2}, Wenying Wei\IEEEauthorrefmark{1}\IEEEauthorrefmark{2}, Xiaofei Xie\IEEEauthorrefmark{3}, Yang Liu\IEEEauthorrefmark{3}, Xiaohong Guan\IEEEauthorrefmark{1}\IEEEauthorrefmark{2}~\IEEEmembership{Fellow, IEEE}, Ting Liu\IEEEauthorrefmark{1}\IEEEauthorrefmark{2}

\IEEEauthorblockA{\IEEEauthorrefmark{1}MOEKLINNS Lab, Department of Computer Science and Technology, Xi'an Jiaotong University, 710049, China}

\IEEEauthorblockA{\IEEEauthorrefmark{2}School of Cyber Science and Engineering, Xi'an Jiaotong University, 710049, China}

\IEEEauthorblockA{\IEEEauthorrefmark{3}School of Computer Science and Engineering, Nanyang Technological University, Singapore}

}

\markboth{IEEE Transactions on Information Forensics and Security,~Vol.~x, No.~x,~x}{xxx
	\MakeLowercase{Fan\textit{ et al.}}: Can We Trust Your Explanations? Sanity Checks for Interpreters in Android Malware Analysis}

\maketitle

\begin{abstract}
With the rapid growth of Android malware, many machine learning-based malware analysis approaches are proposed to mitigate the severe phenomenon. However, such  classifiers are opaque, non-intuitive, and difficult for analysts to understand the inner decision reason. For this reason, a variety of explanation approaches are proposed to interpret predictions by providing important features. Unfortunately, the explanation results obtained in the malware analysis domain cannot achieve a consensus in general, which makes the analysts confused about whether they can trust such results. In this work, we  propose principled guidelines to assess the quality of five explanation approaches by designing three critical quantitative metrics to measure their stability, robustness, and effectiveness. Furthermore, we collect five widely-used malware datasets and apply the explanation approaches on them in two tasks, including malware detection and familial identification. Based on the generated explanation results, we conduct a sanity check of such explanation approaches in terms of the three metrics. The results demonstrate that our metrics can   assess the explanation approaches and help us obtain the knowledge of most typical malicious behaviors for malware analysis. 
\end{abstract}


\section{Introduction}
\label{sec_introduction}

The rapid growth of Android malware has posed significant threats to smartphone users~\cite{zhou2012dissecting}. To mitigate the severe phenomenon, many malware analysis approaches are proposed in recent years, especially the machine learning-based approaches. They first train a classifier using a labeled dataset and then use it for two main tasks, i.e., malware detection (detecting whether a given sample is malicious or benign)~\cite{fan2017dapasa, hou2017hindroid} and familial identification (identifying which family a detected malware belongs to)~\cite{fan2016faldroid, feng2017automated}. The experimental results conducted by such work demonstrate that they can achieve fairly good performance.

However, the constructed classifier model seems like a black-box to the security analysts, since it cannot provide clear evidence to explain why the given sample is identified as a malicious sample or not. Note that it is difficult for analysts to manually analyze the black-box model and infer the decision reason because the size of the training data and the complexity of the learned model are too big for humans to understand~\cite{guidotti2018survey}. Furthermore, existing approaches reveal that the learned models can be easily attacked by adversarial samples~\cite{grosse2017adversarial}, which would even increase the difficulty of malware analysis. As a result,  the analysts cannot determine whether they could trust the decisions, and they are hesitated to  adopt the theoretical approaches into practical applications. 

\begin{figure}[!t]
	\centering
	\includegraphics[width=0.5\textwidth]{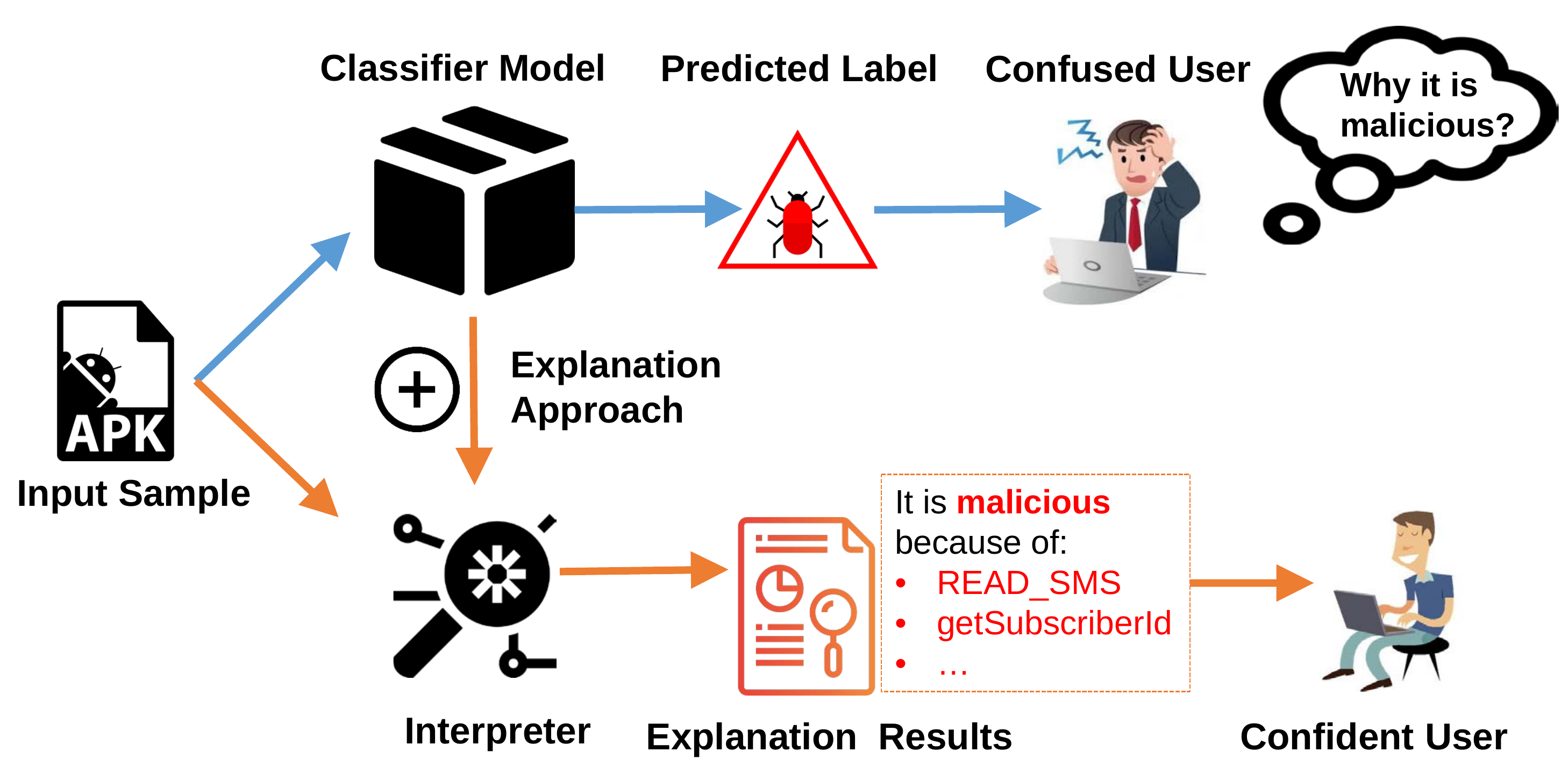}
	\caption{Illustration of the traditional machine learning process and the  interpretable machine learning process. }
	\label{Fig-Flow}
	\vspace{-10pt}
\end{figure}

In order to make users trust the machine learning-based approaches, a variety of explanation approaches have been developed recently to interpret predictions by providing important features. Fig. \ref{Fig-Flow} presents two different decision pipelines between the traditional classifier and constructed interpreter.  For the former pipeline, the user is confused about why the input sample is detected as malicious. For the latter pipeline, after the applying of explanation approach on the classifier model, the interpreter is capable of providing specific reasons for particular machine decisions. 

 Most prominent of these explanation approaches include local, model-agnostic ones~\cite{ribeiro2016should, ribeiro2018anchors, guidotti2018local, lundberg2017unified, guo2018lemna} that focus on explaining individual predictions for a given black-box classifier. The basic idea of these explanation approaches is to approximate the local decision boundary using a linear model to infer the important features of the current input instance. Such explanation features can be leveraged by analysts to digest the security analysis results~\cite{guidotti2019survey}, debug pre-trained models~\cite{krause2016interacting}, and detect adversarial input~\cite{liu2018adversarial}. The developed explanation approaches seem like the keys to open the black-box model and infer the mechanism of decision-making.

  Unfortunately, we find that the explanation results provided by existing explanation approaches cannot achieve a consensus in general (we illustrate a motivation example in Section \ref{sec_motivation} for details), which makes the analysts confused about whether they can trust the explanation results.
  Meanwhile, to the best of our knowledge, there is only a little reliability analysis work of these explanation approaches~\cite{adebayo2018sanity,yeh2019fidelity,oana2019can}. Adebayo \textit{et al.}~\cite{adebayo2018sanity} proposed an actionable methodology based on randomization tests to evaluate the adequacy of explanation approaches. Their results find that some widely deployed saliency approaches are independent of both the data the model was trained on and the model parameters. Yeh \textit{et al.}~\cite{yeh2019fidelity} proposed two objective evaluation metrics, naturally termed infidelity and sensitivity, for machine
  learning explanations.
  Camburu \textit{et al.}~\cite{oana2019can}  introduced an off-the-shelf evaluation test for post-hoc explanation approaches under the feature-selection perspective.    However, there are three main limitations to directly apply these approaches to the Android malware analysis domain. First, 
  the work~\cite{yeh2019fidelity, adebayo2018sanity} mainly focus on the  evaluation of explanation approaches that are designed for the CNNs in the image classification domain. Such approaches require the gradient information of the white-box classifier, and their generalization ability is limited. Second, the work~\cite{oana2019can} proposes two specific metrics that require the ground truth labeled by humans, which is not suitable for our work. Third, the existing proposed metrics are general evaluation properties, while in this work, we need more critical metrics that are available in all cases and in a reasonable time.


	To evaluate the reliability problem of explanation approaches in the critical domain of Android malware analysis, in this paper, we investigate their availability by making the following contributions:
\begin{itemize}
	\item We propose principled guidelines to assess the quality
	of the explanation approaches by designing three quantitative
	metrics to measure their stability, robustness, and
	effectiveness, which are fundamental properties that an explanation approach should satisfy for critical security tasks. 
	
	\item We apply the explanation approaches in Android malware
	analysis, including malware detection and familial identification,
	and conduct a sanity check of such approaches
	in terms of the three proposed metrics. To the best of
	our knowledge, this study is the first work attempting to
	provide a systematized investigation about the availability
	of explanation approaches on malware analysis.
\end{itemize}
  
 The remainder of this paper is structured as follows: Section \ref{sec_motivation} provides the motivating scenario and the explanation approaches evaluated in our work. Section \ref{sec_method} details the data collection and the three proposed metrics. Section \ref{sec_study} reports our study results. After discussing the results and the threats to validity in Section \ref{sec_validity}, we introduce the related work in Section \ref{sec_relatedwork} and concludes this paper in Section \ref{sec_conclusion}.  

\section{Motivation and Explanation Approaches}
\label{sec_motivation}

\subsection{Motivating Scenario}
\begin{figure}[!t]
	\centering
	\includegraphics[width=0.5\textwidth]{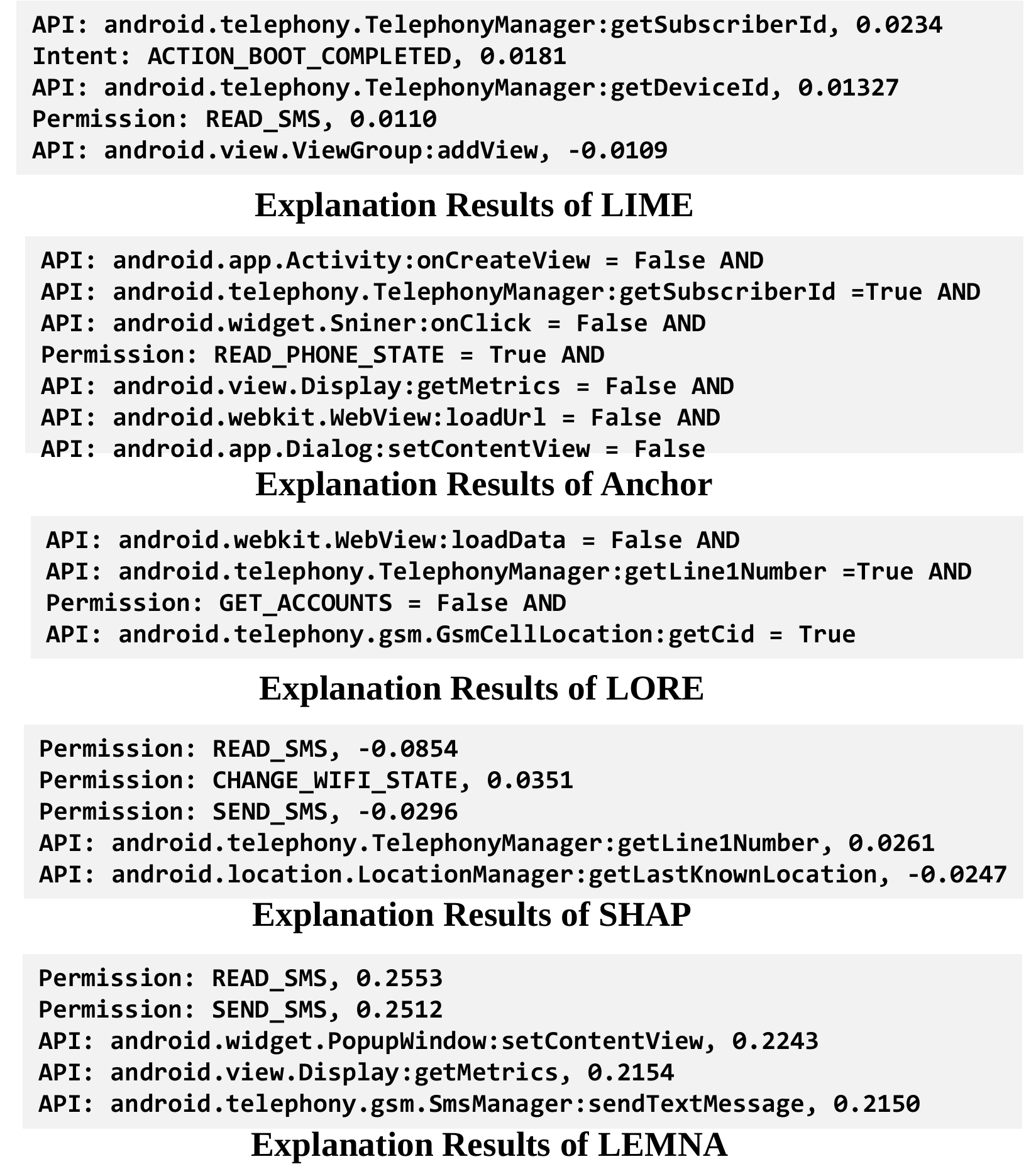}\\
	\caption{Five explanation results for a same app with the same classifier using LIME, Anchor, LORE, SHAP, and LEMNA.}
	\label{Fig-Motivation}
	\vspace{-10pt}
\end{figure}

Let us consider a security analyst who detects malware samples by using a machine learning-based approach. Based on the trained classifier, he feeds the new coming samples to the classifier and gets predicted labels, i.e., malicious or benign. However, the black-box classifier model lacks an explanation for the decision-making. Thus, he tries to leverage the existing explanation approaches to obtain important features of the current input sample. Fig. \ref{Fig-Motivation} presents the explanation results for an app that is predicted as a malware by the same classifier using five different local and model-agnostic explanation approaches, i.e.,  LIME~\cite{ribeiro2016should}, Anchor~\cite{ribeiro2018anchors}, LORE~\cite{guidotti2018local}, SHAP~\cite{lundberg2017unified}, and LEMNA~\cite{guo2018lemna}. The descriptions of these approaches will be introduced later. Note that both LIME, SHAP, and LEMNA provide a parameter to control the number of output explanation features. However, for Anchor and LORE, their explanation results are represented as rules, and their feature numbers differ with inputs rather than controlled by users.  

%
%
%

In Fig. \ref{Fig-Motivation}, the explanation results of LIME, SHAP, and LEMNA contain five features when the parameter value is set as five by the analyst. However, the explanation results of  Anchor and LORE contain seven  and four rules, respectively.  Note that the numbers behind the explanation features in Fig. \ref{Fig-Motivation} denote their corresponding weights. We observe that the explanation results for the five approaches are quite different. For example, there is no common feature for the results of Anchor and LORE. Moreover, for LIME, the feature \texttt{READ\_SMS} shows a positive effect on the prediction of a malicious label, but for SHAP, the same feature shows a negative effect on the same label prediction. 

The above observation makes the analyst confused. He would wonder  which explanation approach he should choose, and can he trust the explanation results? Since there is little standard criterion for measuring explanation approaches in Android malware analysis, such a scenario strongly motivates us to conduct an evaluation study on existing explanation approaches about their availability in this critical domain.

\subsection{Explanation Approaches}
The descriptions of the above explanation approaches are listed below:

\textbf{LIME} is proposed by Ribeiro \textit{et. al}~\cite{ribeiro2016should}, of which 
the intuition is that the explanation can be derived locally from the samples generated randomly in the neighborhood of the input sample, and weighted according to their proximity to it. Specifically, for a given sample $x$, it first creates a set of perturbations of $x$ by  setting some values of $x$ randomly. The set is denoted as $D(x)=\{x'_1, \dots, x'_t \}$.  Based on the given model, it can obtain the corresponding labels of each perturbation sample. Then, it approximates the decision boundary using the model below:
\begin{equation}
\mathop{\arg\min}_{g\in G}\sum^t_{i=1}{dif(x,x'_i)(f(x'_i)-g(x'_i))^2+ \Omega(g)}
\label{eq-LIME}
\end{equation}
where $G$ denotes the set of all linear function and $dif(x,x'_i)$ indicates the difference between the given sample $x$ and a perturbation sample $x'$. $\Omega(g)$ is a penalty term that measures the complexity of explanation results, and it can be the number of non-zero for linear models. Finally, LIME calculates the weights to denote the feature importance by the Lasso algorithm~\cite{efron2004least}.

\textbf{Anchor} is also proposed by Ribeiro \textit{et. al}~\cite{ribeiro2018anchors}, which extracts if-then rules as local interpretations. Anchor uses a bandit algorithm that randomly constructs the anchors with the highest coverage and precision threshold, of which an anchor explanation is a decision rule that sufficiently ties a prediction locally such that changes to the rest of the feature values do not matter. Specifically,  the approach first creates a rule $A(x)$  (a set of feature predicates) that is a sufficient condition for $f(x)$ with high probability. Given a sample $x$ and its perturbation set $D(x)$, $A$ is defined as an anchor if 
\begin{equation}
prec(A)=\mathbb{E}_{D(x'|A)}[\mathbbm{1}_{f(x)=f(x')}]\ge \pi
\label{eq-Anchor}
\end{equation}
where $D(x'|A)$ denotes the conditional distribution when the rule $A$ applies. $\pi$ is the desired level of precision. Since there might exist multiple anchors that are able to meet this criterion, the approach selects the anchor with the largest coverage of perturbation samples as the final explanation for $x$. 

\textbf{LORE} is proposed by Guidoti \textit{et. al}~\cite{guidotti2018local}, which takes the similar way to interpret the machine learning models like Anchor. However, there are two main differences. First, LORE generates perturbation samples in the vicinity of  $x$ using 
genetic algorithm. The perturbation samples  are composed of two parts, the set $D_{=}=\{x'|f(x)=f(x')\}$ and the set $D_{\neq}=\{x'|f(x)\neq f(x')\}$. 
The algorithm calculates the two sets by maximizing the following two fitness functions:
\begin{equation}
fitness^x_{=}(x')=I_{f(x)=f(x')}+(1-dif(x,x'))-I_{x=x'}
\label{eq-LORE-1}
\end{equation}
\begin{equation}
fitness^x_{\neq}(x')=I_{f(x)\neq f(x')}+(1-dif(x,x'))-I_{x=x'}
\label{eq-LORE-2}
\end{equation}
where $I_{true}=1$ and $I_{false}=0$. Then, LORE trains a decision tree using the two sets to extract a logic rule to explain why $x$ is labeled as a specific class.

\textbf{SHAP}  is proposed by Lundberg and Lee~\cite{lundberg2017unified}, which focuses on local linear interpretable methods. SHAP follows a similar approach as LIME but uses the SHAP kernel as a weighting function, which creates SHAP values when solving the regression. Specifically, for LIME, the $\Omega(g)$ is the number of non-zero weights in the linear model, and $dif(x,x')$ is obtained based on cosine distance. SHAP grounds these definitions in game-theoretic principles to guarantee that the
explanations satisfy certain desired properties, including local accuracy, missingness, and consistency.

\textbf{LEMNA} is proposed by Guo \textit{et. al}~\cite{guo2018lemna}, which is designed to handle feature dependency and handle nonlinear local boundaries. It leverages a mixture regression model for approximation, which is a weighted sum of $M$ linear models:
\begin{equation}
f(x)=\sum^M_{j=1}{\theta_j(\beta_j x+\epsilon_j)}
\label{eq-LEMNA}
\end{equation}
where $M$ denotes the model numbers, and $\theta$ denotes the weights for each model. The variables $\epsilon$ originate from a normal distribution. The variables $\beta_j$ are the regression coefficients that can be interpreted as $M$ linear approximations. 

In addition to the above five explanation approaches, there are many other ones, such as input perturbation based approaches and back propagation based approaches, which will be introduced in the related work. 

In this work, we conduct our study on the above five post-hoc explanation approaches because of two main reasons. First, they are model-agnostic approaches. Unlike the other explanation approaches that can only be applied in specific models such as CNN or GNN, the five approaches studied in this work are more suitable for the malware analysis domain since most malware analysis approaches still rely on conventional machine learning algorithms. Second, they are local interpretability methods, which are  suitable for explaining the decision-making of why a target sample is malicious. Combining the two requirements, we select the five approaches.

\subsection{Basic Notations}
In this section, we introduce a set of basic notations used in our paper, which are listed in Table \ref{Tab-Notation}, as well as their definitions. The notations are categorized into three groups listed below.

\begin{table}[t]
	\newcommand{\tabincell}[2]{\begin{tabular}{@{}#1@{}}#2\end{tabular}}
	\centering
	\scriptsize
	\caption{Notations and definitions}
	\begin{tabular}{cc}
		\toprule[1.5pt]
		Notation & Definition \\ \hline
		$\mathbf{x_i}\in \mathbb{R}^d$ &  the $d$-dimensional feature vector of the $i$-th sample               \\ \hline	
		$y_i, \hat{y_i} \in \mathcal{C}$&     \tabincell{c}{true label and predicted label \\ of the $i$-th  sample,  $\mathcal{C}$ denotes the label set}       \\ \hline
		$f$      &   a classifier model constructed on a training dataset       \\ \hline
		$m$    &   an explanation approach \\ \hline
		$g=m(f)$      &   \tabincell{c}{specific interpreter constructed based on\\ an explanation method $m$ and a trained classifier $f$ } \\ \hline
		$e_i(g)$ & \tabincell{c}{the explanation results of the $i$-th sample with  interpreter $g$}  \\ \hline
		$stb(m, \mathcal{T})$ &  \tabincell{c}{stability of explanation approach  $m$ on testing dataset $\mathcal{T}$ }\\ \hline
		$rob(m, \mathcal{T})$ & \tabincell{c}{robustness of explanation approach $m$ on testing dataset $\mathcal{T}$ }\\ \hline
		$eff(m, \mathcal{T})$ & \tabincell{c}{effectiveness of explanation approach $m$ on testing dataset $\mathcal{T}$ }\\ \hline
	\end{tabular}
	\label{Tab-Notation}
	\vspace{-10pt}
\end{table}

\vspace{2mm}
\noindent\textbf{Malware Classifier:} Given a set of training samples, each sample $x_i$ is represented as a $d$-dimensional feature vector $\mathbf{x_i}\in  \mathbb{R}^d$, a classifier $f$ can be constructed on the dataset with a specific machine learning algorithm. We use $f(\mathbf{x_i})=\hat{y_i}$ to denote the prediction process of the $i$-th sample. The classifier would conduct a correct classification if the predicted label $\hat{y_i}$ equals to the true label $y_i$. Note that for malware detection task, the label set $\mathcal{C}$ contains only two elements, i.e., benign and malicious. However, for the familial identification task, the element in $\mathcal{C}$ denotes the family name, such as \textit{adrd}, \textit{genimi}, and \textit{droidkungfu}.

\vspace{2mm}
\noindent\textbf{Interpreter:} An interpreter $g=m(f)$ here is generated by applying an explanation approach $m$ on a specific pre-trained classifier $f$. Given a test sample  $x_i$, the interpreter $g$ will output its explanation result $e_i(g)$, which is represented as a sorted set of features that are ranked based on their importance  to the decision result. 

\vspace{2mm}
\noindent\textbf{Quantitative Metrics:} For convenience we use $stb(m, \mathcal{T})$, $rob(m, \mathcal{T})$, and $eff(m, \mathcal{T})$ to denote the stability, robustness, and effectiveness of an explanation approach $m$ on a testing dataset $\mathcal{T}$, respectively. The values of stability and effectiveness  range from 0 to 1, while the value of robustness ranges from -1 to 1. Here a higher value indicates a  better explainability. The detail calculation of the three metrics will be introduced in Section \ref{sec_method}.

\section{Study Design}
\label{sec_method}

In study design, we first introduce the data collection procedure that contains the construction of datasets used for malware detection and familial identification, and the extraction of useful features from the APK files. Then we detail the intuitions of the three proposed quantitative metrics, as well as their calculation equations.

\subsection{Data Collection}
\label{subsec_data}

\noindent\textbf{Dataset Construction:} Five widely-used malware datasets are evaluated in our work. They are provided by Genome project~\cite{zhou2012dissecting}, Drebin~\cite{arp2014drebin}, FalDroid~\cite{fan2018android}, RmvDroid~\cite{wang2019rmvdroid} and AMD~\cite{wei2017deep}\footnote{The original AMD dataset contains 24,650 samples, which is a time-consuming job to calculate the explanation results for all samples. Therefore, we randomly select one-fifth samples from each family.}.  For convenience, they are named as dataset-I, dataset-II, dataset-III, dataset-IV and dataset-V. Their descriptions are listed in Table \ref{Tab-MalwareDataset}, where columns 2-3 list the number of families and the number of malware samples, columns 4-6 list the maximum, minimum, and average size of samples, and column 7 lists the producing time of the corresponding samples. Note that the authors of dataset-V select three samples as representatives from each malware family to conduct in-depth manual analysis. Therefore, their analysis result can be regarded as the ground truth to check the accuracy of the explanation results.

 Each sample has been attached to a family label given by experts, and they are widely-used as the ground truth for malware analysis. For the dataset-I provided by Genome project, all the samples were manually analyzed by Jiang and Zhou~\cite{zhou2012dissecting}, so well as their family labels. Specifically, they labeled the malware samples according to their malicious activities. For example, the samples in \textit{GingerMaster} family generally get root permissions, steal sensitive information, and then send it to a remote server. 
For the other four datasets, their malware samples are labeled based on VirusTotal~\cite{virustotal}, which is a system with more than 50 anti-virus scanners (e.g., AVL, McAfee, and ESET-NOD32). Given a malware sample, each anti-virus scanner will return a family label. However, there are two issues for these anti-virus scanners. First, the family labels given by different scanners are not always the same (e.g., \textit{Plankton/Plangton/planktonc}). Second, the results of the scanners rarely reach a consensus. To address these issues, the authors of the four datasets initially constructed a family label dictionary. Then, they labeled the malware with the family name that is agreed by more than half of the scanners.

For benign apps, we construct a benign dataset that contains 10,000 samples that are collected from Google Play~\cite{Google}. Each sample has been uploaded to the VirusTotal~\cite{virustotal} to ensure that all scanners report it as benign.

There are two malware analysis tasks, i.e., malware detection and familial identification. The former is a binary classification problem, and the latter is a multi-class classification problem. For malware detection, we first mix each malware dataset with the benign dataset and obtain five mixed datasets. Then we randomly split each mixed dataset into two parts, one is used for training, and the other one is used for testing. For familial identification, half of the samples in each family are used for training, and the remaining samples are used for testing. The constructed datasets used for malware detection and familial identification are listed in Table \ref{Tab-TrainTest}. Here the training size and testing size are not equal because if a family contains only one sample, we put it in the training dataset, and if the sample size of a family is odd, we put the extra sample in the testing dataset. 

\begin{table}[!t]
	\centering
	\scriptsize
	\caption{Descriptions of five widely-used malware datasets.}
	\scalebox{1.0}{
		\begin{tabular}{p{1.5cm}<{\centering}p{0.5cm}<{\centering}p{0.7cm}<{\centering}p{0.5cm}<{\centering}p{0.5cm}<{\centering}p{0.5cm}<{\centering}p{1.5cm}<{\centering}}
			\toprule[1.5pt]
			Dataset  & \#Family &  \#Malware  &  Max. &  Min. &  Avg. & Time \\ \midrule
			dataset-I~\cite{zhou2012dissecting} & 49  &   1,260 & 15.4MB & 12KB & 1.3MB  & 2011--2012 \\ %
			dataset-II~\cite{arp2014drebin}  &  179  &  5,560  & 24.8MB  & 5KB & 1.3MB  & 2011--2014 \\ %
			dataset-III~\cite{fan2018android} &  36  &   8,407  & 36.2MB & 12KB & 2MB & 2013--2014 \\
			dataset-VI~\cite{wang2019rmvdroid}  & 56 & 9,133 &  76.1MB &  48KB  &   4.8MB &  2014--2018  \\
			dataset-V~\cite{wei2017deep} & 71 & 4,741 & 47.2MB &   9.8KB   &  2.1MB   &  2010--2016 \\
			\bottomrule[1.5pt]
		\end{tabular}
	}
	\label{Tab-MalwareDataset}
		\vspace{-10pt}
\end{table}

\begin{table}[!t]
	\centering
	\scriptsize
	\caption{Descriptions of the datasets used for malware detection (MD) and familial identification (FI).}
	\scalebox{1.0}{
		\begin{tabular}{cccccc}  \toprule[1.5pt]
			\multirow{2}{*}{Dataset} & \multicolumn{2}{c}{Training} & \multicolumn{2}{c}{Testing} & \multirow{2}{*}{\#Label} \\  
			& \#Malware     & \#Benign     & \#Malware     & \#Benign    &                          \\ \hline
			MD-I                     & 633           & 5,000         & 627           & 5,000        & 2                        \\
			MD-II                    & 2,774          & 5000         & 2,786          & 5,000        & 2                        \\
			MD-III                   & 4,194          & 5,000         & 4,213          & 5,000        & 2                        \\
			MD-IV                    & 4,554          & 5,000         & 4,579          & 5,000        & 2                        \\
			MD-V                     &  2,353             &  5,000        &  2,388        &  5,000     & 2      \\
			FI-I                     & 633           & -            & 627           & -           & 49                       \\
			FI-II                    & 2,774          & -            & 2,786          & -           & 179                      \\
			FI-III                   & 4,194          & -            & 4,213          & -           & 36                       \\
			FI-IV                    & 4,554          & -            & 4,579          & -           & 56                      
			\\ 
			FI-V                   &    2,355          &   -      &      2,388       &     -      &    71  \\ \bottomrule[1.5pt]
		\end{tabular}
	}
	\label{Tab-TrainTest}
\vspace{-20pt}
\end{table}

\vspace{2mm}
\noindent\textbf{Feature Extraction:} Unlike the inartificial word features used for text classification and the  pixel features used for image recognition, the features used for malware analysis are manually designed by researchers. Note that in this work, our goal is not to propose a novel malware analysis feature that achieves better performance than state-of-the-art approaches. We aim to conduct an assessment of the quality of different explanation approaches in Android malware analysis. Thus, we rely on the 295 features provided by FeatureSmith~\cite{zhu2016featuresmith}, which consists of 259 application
programming interface (API) calls, 33 permissions\footnote{Android permission control is one of the major Android security mechanisms. Android permissions are requested by apps before the apps can use
	certain system data and features.}, and 3 intents\footnote{An intent is a bundle of information describing a desired action, including
	the data to be acted upon, the category of a component that should perform the
	action, and other pertinent instructions.}. Note that such features are collected in published knowledge, e.g., scientific papers, and each one contains a set of   related informative behaviors, based on which we can provide more comprehensible explanations. For example, the API \texttt{getSimOperatorName} is related to the semantic behaviors of ``return privacy-sensitive information'' and ``leak to remote web server''. 

To construct the feature vectors for each app, we first leverage the mature disassembling tools, such as \texttt{apktool}~\cite{Apktool}, to obtain the \texttt{AndroidManifest.xml} file and the Dalvik code from the APK files. The  \texttt{AndroidManifest.xml} file contains essential information about an app to the Android system, including the requested permissions and intents. Moreover, we can identify the used API calls by scanning the Dalvik code files. It is worth noting that the widely-used third-party and advertisement libraries might affect the performance of malware analysis, so well as the explanation results. We filter out these libraries from the Dalvik code with the blacklist provided by existing approaches~\cite{li2017libd, li2016investigation}. For each feature, if an app contains it, then the corresponding feature value is set as 1; otherwise, it is set as 0. Finally, for malware detection, we attach the labels ``malicious'' and ``benign'' to their constructed feature vectors, and for familial identification, we attach the family labels to the corresponding malware samples.

\subsection{Intuitions of Quantitative Metrics}

Before introducing the measurement of the metrics, we list their importance and intuitions:

\vspace{2mm}
\noindent\textbf{Stability}: The explanation results generated in a critical security system should be stable. That is, the explanation features must not be seriously influenced by the fluctuation techniques used in the explanation approaches. As introduced before, the local explanation approaches interpret individual predictions of any given classifier by learning an interpreter (e.g., linear model) locally around each prediction. Note that the learned interpreter is trained with a set of generated similar instances around the given sample; however, since the distribution of generated instances is not fixed, which would cause instability for the explanation result. Considering that if the generated explanations vary a lot between multiple runs for the same input, the interpreter user would be quite confused and not trust the explanation results. In other words, if an explanation approach can really capture the actual reason for an individual prediction, then the explanation results should remain the same on similar pre-trained models. Thus, we regard the stability as the first fundamental property of local explanation approach. 

\vspace{2mm}
\noindent\textbf{Robustness}: Having an explanation result for individual prediction is necessary but not enough. The explanation itself should also be robust in order to build human trust. Existing works~\cite{ghorbani2019interpretation,heo2019fooling}  have demonstrated that the explanation results might be altered by small systematic perturbations to the input data while preserving the predicted label. Taking the pathology diagnosis task as an example, an explanation approach would suggest that a particular section in an image is important for the tumor prediction. It would be highly disconcerting that, if in a remarkably similar image, a very different section is regarded as the explanation result while the predicted label remains the same. Weak robustness of explanation approaches would cause huge damage in critical sceneries. Therefore, we regard the robustness as another fundamental property of an explanation approach, which is used to measure how similar the explanation results are for similar instances.

\vspace{2mm}
\noindent\textbf{Effectiveness}: The primary goal of an explanation approach is to find the decision basis for an individual prediction. If an explanation approach just returns several carefully picked features that have no relation to the prediction result, it can easily achieve high stability and robustness scores. However, such returned features are useless to interpreter users. For example, in the image classification task, an edge detector is a classical tool to highlight sharp transitions in an image. It is untrained and does not depend on any predictive model. It can just be regarded as a function of a given input image. As a result, if the edge detector is used as an explanation approach, it can achieve very high stability and robustness scores since the output results remain the same~\cite{adebayo2018sanity}. To address the above problem, we regard the effectiveness as the third fundamental property, to measure whether the explanation results are important to the decision-making. In other words, if the explanation results are really the decision basis for an individual prediction, then after the removal of such features, the prediction result would be changed.

\subsection{Measurement of Quantitative Metrics}
\label{subsec_metric}

\noindent\textbf{Similarity calculation between explanations:}
To measure the three quantitative metrics, we need to define the similarity calculation between explanation results that are represented as feature sets. 
Considering that the sizes of explanation features for rule-based methods, i.e., Anchor and LORE, are not fixed, here we calculate the explanation similarity based on top-$k$ intersection of the generated feature sets.  
Specifically, given two explanation results, $e_i(g)$ and $e_j( g')$, and a parameter $k$, their similarity is obtained based on the dice coefficient as Eq. (\ref{eq-dice}),
\begin{equation}
\small{
	sim(e_i(g), e_j(g'), k)=2*\frac{e_i^k(g)\cap e_j^k( g')}{|e_i^k(g)|+|e_j^k(g')|}
}
\label{eq-dice}
\end{equation}
where $e_i^k(g)$ denotes the top-$k$ features of the $i$-th sample's explanation result generated by interpreter $g$. Note that, for Anchor and LORE, if the feature number of their explanation results is less than $k$, then we would use all their explanation features. For example,  if $k$ is set as 4, the similarity between the explanation results of LORE and SHAP in Fig. \ref{Fig-Motivation} is calculated based on their top-4 features, i.e., $2*1/(4+4)=0.25$. However, if $k$ is set as 5, which is bigger than the feature number of LORE result, we use all its 4 explanation features and the top-5 features of SHAP. Therefore, the explanation similarity between LORE and SHAP is $2*1/(4+5)=0.22$. 


\vspace{2mm}
\noindent\textbf{Stability:}
In our work, the stability of an explanation approach $m$, denoted as $stb(m, \mathcal{T})$, is measured on a target testing dataset $\mathcal{T}$. Specifically, it is obtained with Eq. (\ref{eq-stbMethod}),
\begin{equation}
\small{
	stb(m, \mathcal{T})=\frac{1}{|\mathcal{T}|}*\sum_{x_i \in \mathcal{T} }{stb(m,x_i)}
}
\label{eq-stbMethod}
\end{equation}
where $stb(m, x_i)$ denotes the stability of explanation results for the sample $x_i$ with $m$ on a set of similar pre-trained models. The similar model set is represented as $F=\{f_p|1\le p\le \alpha\}$, where $\alpha$ denotes the number of pre-trained models and it is bigger than two. Note that the models in $F$ are trained with very similar arguments and they present  approximate performance. Based on $F$, we can obtain a set of similar interpreters, represented as $G=\{g_p=m(f_p)|1\le p \le \alpha \}$.

Then $stb(m, x_i)$ is obtained by calculating the average dice coefficient similarity between the explanation results generated by similar interpreters (i.e., $g_p, g_q \in G$) as Eq. (\ref{eq-stbSample}). 
\begin{equation}
\small{
	stb(m, x_i)=\frac{1}{\binom{\alpha}{2}}*\sum_{g_p, g_q\in G, p\neq q}{sim(e_i(g_p), e_i(g_q), k)}
}
\label{eq-stbSample}
\end{equation}

%




\vspace{2mm}
\noindent\textbf{Robustness:}
When measuring the robustness for malware analysis task, the samples in the same families can be regarded as the similar samples since they conduct same malicious behaviors. Intuitively, for familial identification, the explanation results of the samples that are predicted in the same family should be highly similar even if there exists a bit difference between their feature vectors. Conversely, the explanation results of the samples in different families should be highly different. Specifically, the robustness of an explanation approach $m$ on testing dataset $\mathcal{T}$ is calculated as Eq. (\ref{eq-robMethod}), 
\begin{equation}
\small{
rob(m, \mathcal{T})=\frac{1}{|\mathcal{T}|}*\sum_{x_i\in \mathcal{T}}{rob(m,x_i)}
}
\label{eq-robMethod}
\end{equation}
where $rob(m,x_i)$ denotes the robustness of explanation results for the sample $x_i$ with $m$. To calculate $rob(m, x_i)$, we first construct two sets, $\mathcal{S}({x_i})$ and $\mathcal{D}({x_i})$, which denote the set of samples with the same predicted label as $x_i$ and the set of samples with the different predicted label as $x_i$, respectively. The samples in the two sets are obtained from $\mathcal{T}$. The two sets are represented as Eq. (\ref{eq-S}) and Eq. (\ref{eq-D}).
\begin{equation}
\small
\mathcal{S}({x_i})=\{x_j| \hat{y_j} = \hat{y_i}\}
\label{eq-S}
\end{equation}
\begin{equation}
\small
\mathcal{D}({x_i})=\{x_t| \hat{y_t} \neq \hat{y_i}\}
\label{eq-D}
\end{equation}

 After that, we calculate the average explanation similarity between $x_i$ and the samples in the two sets. They are represented as $avg(x_i, \mathcal{S}({x_i}))$ and $avg(x_i, \mathcal{D}({x_i}))$, and are obtained with Eq. (\ref{eq-avgS}) and Eq. (\ref{eq-avgD}). Note that here there is only one fixed interpreter here, thus we use $e_i$ instead of $e_i(g)$  for convenience.
\begin{equation}
\small
avg(x_i, \mathcal{S}({x_i}))=\frac{1}{|\mathcal{S}({x_i})|}*\sum_{x_j\in \mathcal{S}({x_i})}{sim(e_i, e_j, k)}
\label{eq-avgS}
\end{equation}
\begin{equation}
\small
avg(x_i, \mathcal{D}({x_i}))=\frac{1}{|\mathcal{D}({x_i})|}*\sum_{x_t\in \mathcal{D}({x_i})}{sim(e_i, e_t, k)}
\label{eq-avgD}
\end{equation}

 Finally, $rob(m, x_i)$ is calculated as Eq. (\ref{eq-robX}) based on the numerical difference between $avg(x_i, \mathcal{S}({x_i}))$ and $avg(x_i, \mathcal{D}({x_i}))$.
\begin{equation}
\small
rob(m,x_i)=avg(x_i, \mathcal{S}({x_i}))-avg(x_i, \mathcal{D}({x_i}))
\label{eq-robX}
\end{equation}


%

\vspace{2mm}
\noindent\textbf{Effectiveness:}
In our work, the effectiveness of an explanation approach $m$ on testing dataset $\mathcal{T}$  is calculated as Eq. (\ref{eq-effMethod}),
\begin{equation}
\small
eff(m, \mathcal{T})=\frac{1}{|\mathcal{T}|}*\sum_{x_i\in \mathcal{T}}{eff(m,x_i)}
\label{eq-effMethod}
\end{equation}
where $eff(m,x_i)$ denotes the effectiveness of explanation results for the sample $x_i$ with $m$.  To calculate $eff(m,x_i)$, we first construct a new feature vector for the given sample $x_i$ by mutating the feature values according to the top-$k$ explanation features as Eq. (\ref{eq-mutate}). For example, if the value of a target feature is 1, and it is listed in the explanation results, then in the new feature vector, its value will be changed to 0. 
\begin{equation}
\small
\mathbf{x_i^*}=mutate(\mathbf{x_i}, e_i^k)
\label{eq-mutate}
\end{equation}

After the mutation of feature vectors, we check whether the prediction result is changed. Based on Eq. (\ref{eq-effSample}) and Eq. (\ref{eq-predict}), if the prediction result is changed, $eff(m,x_i)$ is assigned with 1, indicating that such features are important to the current decision-making. Otherwise,  $eff(m,x_i)$ is 0. 

\begin{equation}
\small
eff(m, x_i)=
\begin{cases}
1, & {\hat{y_i^*}\neq \hat{y_i}} \\
0, & {\hat{y_i^*}=\hat{y_i}}
\end{cases}
\label{eq-effSample}
\end{equation}
\begin{equation}
\small
\hat{y_i^*}=f(\mathbf{x_i^*})
\label{eq-predict}
\end{equation}

\begin{table}[!t]
	\centering
	\scriptsize
	\caption{Performance of malware detection on five datasets using different classifiers.}
	\scalebox{1.0}{
		\begin{tabular}{ccccccc}  \toprule[1.5pt]
			Dataset                 & Classifer & TPR & FPR & Precision & Recall & F-measure \\ \hline
			\multirow{5}{*}{MD-I}   & MLP       &   0.9665  &  0.0041   &  0.9665      &  0.9665      &   0.9665        \\
			& RF        &  0.9585    &  0.0004   &    0.9967       &    0.9585    &    \textbf{0.9772}       \\
			& KNN    &   0.9059  &    0.0063   &   0.9466     &     0.9059    &    0.9258   \\
			& SVM       &  0.9442   &   0.0037  &    0.9689       &   0.9442     &  0.9564         \\ \hline
			\multirow{5}{*}{MD-II}  & MLP       &  0.9784   &  0.0249   &    0.9558       &  0.9784      &   0.9670        \\
			& RF        &  0.9720    &  0.0128   &    0.9765       &  0.9720      &   \textbf{0.9742}        \\
			& KNN       &   0.9407  &    0.0314  &   0.9428    &   0.9407    &   0.9417     \\ 
			& SVM       &   0.9440  &  0.0241   &    0.9556       &   0.9440     &   0.9498        \\ \hline
			\multirow{5}{*}{MD-III} & MLP       &  0.9869   &   0.0304  &   0.9642   &   0.9869     &   0.9754        \\
			& RF        &  0.9802   &  0.0217   &   0.9740        &  0.9802      &   \textbf{0.9771}        \\
			& KNN       &  0.9553   &   0.0399   &    0.9522    &   0.9553    &    0.9537      \\
			& SVM       &   0.9703  &  0.0330   &    0.9607       & 0.9703    &   0.9655        \\ \hline
			\multirow{5}{*}{MD-IV}  & MLP       &   0.9757   &  0.0336   &   0.9633        &    0.9757    &    0.9695       \\
			& RF        &  0.9777   &  0.0316   &    0.9654       &  0.9777      &   \textbf{0.9715}        \\
			& KNN       & 0.9495    &   0.0604  &   0.9342      &     0.9495     &     0.9418     \\
			& SVM       &  0.9600   &  0.0557   &    0.9397       & 0.9600       &   0.9497       \\ \hline
			
				\multirow{5}{*}{MD-V}  & MLP       &   0.9577   &  0.0124   &   0.9731        &    0.9577    &    \textbf{0.9653}       \\
			& RF        &  0.9631   &  0.0162   &    0.9656      &  0.9631      &   0.9644        \\
			& KNN       & 0.9195    &   0.0335  &   0.9281      &     0.9195     &     0.9238     \\
			& SVM       &  0.9095   &  0.0320   &    0.9305       & 0.9095       &   0.9199       \\ \bottomrule[1.5pt]

		\end{tabular}
	}
	\label{Tab-MDPerformance}
	\vspace{-2pt}
\end{table}

\begin{table}[!t]
	\centering
	\scriptsize
	\caption{Performance of familial identification on five datasets using different classifiers.}
	\scalebox{1.0}{
		\begin{tabular}{ccc|ccc} \toprule[1.5pt]
			Dataset                & Classifer & Accuracy & Dataset                 & Classifer & Accuracy \\ \hline
			\multirow{4}{*}{FI-I}  & MLP       &     0.9585     & \multirow{4}{*}{FI-II} & MLP       &     0.9310   \\
			& RF        &     \textbf{0.9760}     &                & RF        &     \textbf{0.9493}     \\
			& KNN      &       0.8692     &                         & KNN      &       0.8668 \\
			
			& SVM       &     0.9649     &                         & SVM       &       0.9454  \\ \hline
			\multirow{4}{*}{FI-III} & MLP       &     0.9064       & \multirow{4}{*}{FI-IV}  & MLP       &     0.7796     \\
			& RF        &    \textbf{0.9150 }     &                         & RF        &   \textbf{0.8142}       \\
			&  KNN      &    0.8609   &                           & KNN       &      0.7770   \\
			& SVM       &    0.9107       &                         & SVM       &     0.7968     \\ \hline
			
			\multirow{4}{*}{FI-V} & MLP       &     0.8846       & \multirow{4}{*}{}  &       &        \\
			& RF        &    \textbf{0.8990 }     &                         &       &         \\
			&  KNN      &    0.8115   &                           &    &        \\
			& SVM       &    0.8944      &                         &        &         \\

			\bottomrule[1.5pt]
		\end{tabular}
	}
	\label{Tab-FIPerformance}
	\vspace{-8pt}
\end{table}

\section{Study Results}
\label{sec_study}

In this section, we first construct the classifier models and present the performance results for malware analysis on five datasets introduced in Section \ref{subsec_exp_model}. Then,  we apply five explanation approaches on the constructed models and obtain the corresponding explanation results, based on which we present the study results by answering RQ1-RQ6 (Section \ref{subsec_exp_inconsistency}-\ref{subsec_exp_effciency}). Specifically,  RQ1 is proposed to prove the significance of our motivation. RQ2-RQ4 are proposed to conduct a quantitative comparison between the introduced five explanation approaches with our three metrics. RQ5 is proposed to  measure the efficiency of the explanation approaches in practice. RQ6 is proposed to measure the generalization ability of our metrics on different classifiers. 


\subsection{Malware Analysis Models}
\label{subsec_exp_model}

Based on the collected dataset and extracted features introduced in Section \ref{subsec_data}, we construct a set of classifiers with multilayer perceptron (MLP), random forest (RF),  k-nearest neighbors (KNN), and support vector machines (SVM). The  classification algorithms are implemented based on sklearn package. The parameters of the classifiers are mainly selected using the default values of sklearn package. Specifically, SVM is trained with the rbf kernel and $1\mathrm{e}{-7}$ tolerance. MLP is trained with 3 layers, where each layer consists of 128 neurons. In addition, the activation function is relu, the solver is adam, and the batch size is 200. RF is trained with 100 trees and a gini criterion. KNN is trained with 10 neighbors, uniform weights, and 30 leaf size.
These algorithms are used in a wide variety of applications by machine learning engineers across the world. 
The performance results for malware analysis, including malware detection and familial identification, are listed in Table \ref{Tab-MDPerformance} and Table \ref{Tab-FIPerformance}, where the numbers in bold denote the best performance in terms of F-measure and accuracy among the four algorithms. 

We can observe that for both malware detection and familial identification, the constructed classifier models can achieve fairly good performance, where all the F-measures are higher than 0.9, and most of the accuracies are higher than 0.8. Furthermore, RF performs best among the four algorithms for most datasets except MD-V, since it is an ensemble classifier that consists of hundreds of trees. However,  the complex tree ensembles are also opaque to us to understand the inner working. 
The main reason is that even the basic decision tree is considered easily understandable and interpretable for humans; however, the final prediction of RF is obtained based on hundreds of basic decision trees, which significantly increases the understanding difficulty of the inner working~\cite{guidotti2019survey}.
Therefore, RF seems like a black-box to us and we take it as a hole to conduct the measurement for existing explanation approaches. 

\begin{table}[!t]
	\centering
	\scriptsize
	\caption{Average feature numbers of explanation results for Anchor and LORE on different datasets.}
	\scalebox{1.0}{
		\begin{tabular}{ccc|ccc}  \toprule[1.5pt]
			Dataset & Anchor & LORE & Dataset & Anchor & LORE \\ \hline 
			MD-I    & 5.4    & 3.4  & FI-I    & 6.7    & 3.2  \\
			MD-II   & 5.4    & 3.6  & FI-II   & 13.4   & 3.7  \\
			MD-III  & 2.9    & 2.9  & FI-III  & 14.3   & 3.7  \\
			MD-IV   & 3.4    & 2.9  & FI-IV   & 17.8   & 3.8 \\ 
			MD-V   & 4.9    & 3.0  & FI-IV   & 17.5   & 4.1 \\                                                      \bottomrule[1.5pt]
		\end{tabular}
	}
	\label{Tab-ExpFeatureNum}
	\vspace{-12pt}
\end{table}

\begin{figure}[!t]
	\centering
	\includegraphics[width=0.4\textwidth]{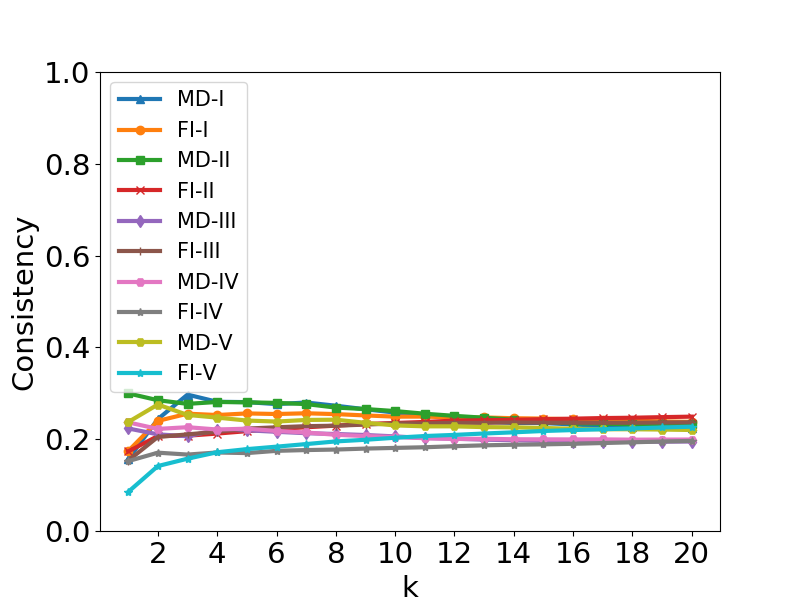}\\
	\caption{Consistency of the explanation results for the samples on different datasets with $k$ from 1 to 20.}
	\label{Fig-RQ1-Inconsistency}
	\vspace{-18pt}
\end{figure}

\begin{figure*}[!t]
	\centering
	\includegraphics[width=\textwidth]{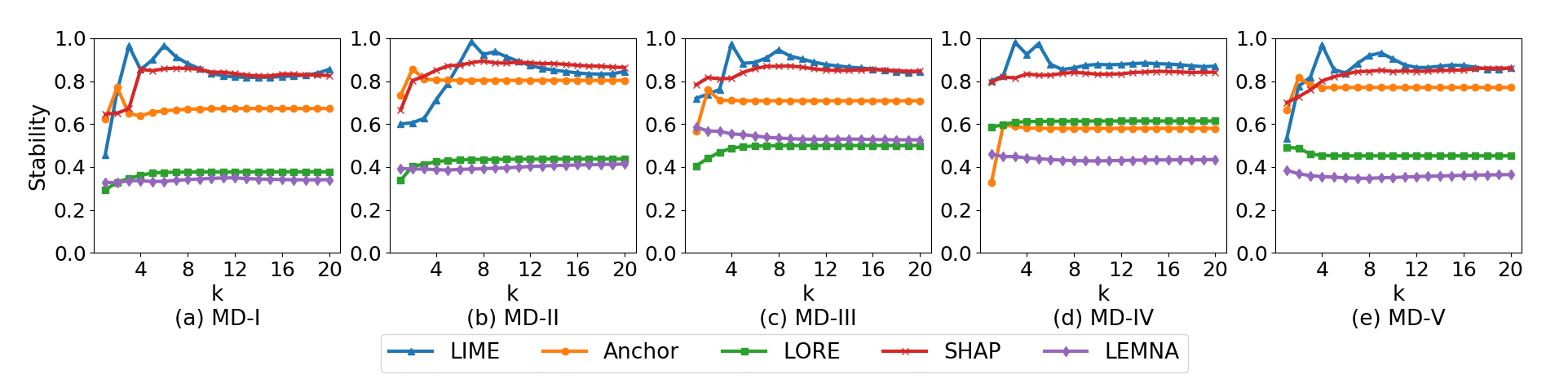}\\
	\caption{Stability of the five explanation  approaches on different malware detection datasets with $k$ from 1 to 20.}
	\label{Fig-RQ2-Stability-MD}
		\vspace{-12pt}
\end{figure*}

\begin{figure*}[!t]
	\centering
	\includegraphics[width=\textwidth]{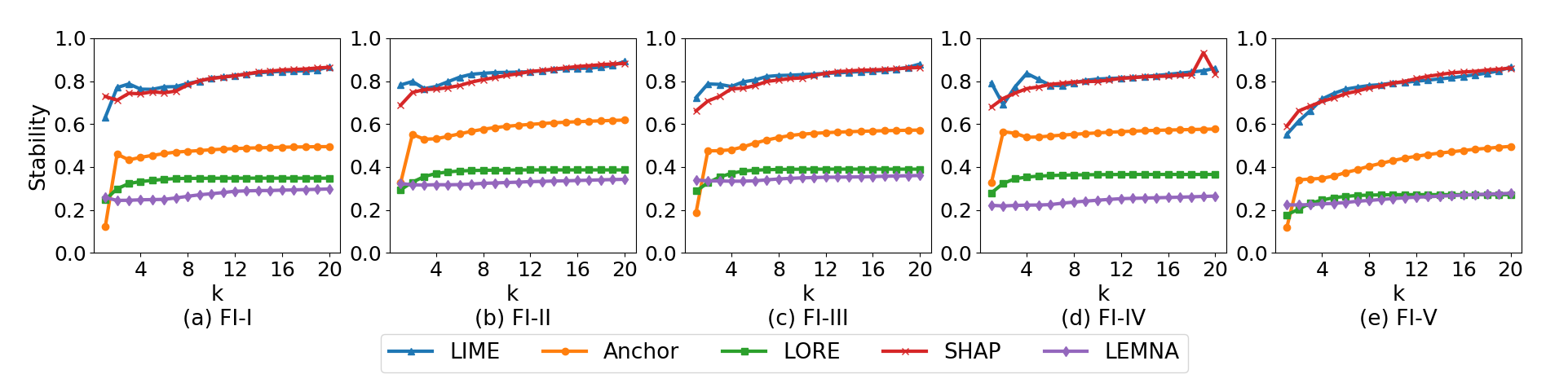}\\
	\caption{Stability of the five explanation  approaches on different familial identification datasets with $k$ from 1 to 20.}
	\label{Fig-RQ2-Stability-FI}
		\vspace{-18pt}
\end{figure*}

%


\subsection{RQ1: To what extent is the inconsistency for the interpretations with different explanation approaches?}
\label{subsec_exp_inconsistency}

To answer RQ1, we apply the five explanation approaches on the constructed RF classifier models and obtain the explanation results for each input sample. For LIME, SHAP, and LEMNA, we set the parameter $k$ from 1 to 20, and obtain their corresponding explanation results. For Anchor and LORE, the average feature numbers of their explanation results on different datasets are presented in Table \ref{Tab-ExpFeatureNum}.  We can observe that the feature numbers of the explanation results generated for familial identification are generally bigger than those for malware detection. The main reason is that familial identification  is a multi-class  classification problem which requires more explanation features to interpret the result, while malware detection is a binary classification problem.

After the generation of explanation results, we calculate the consistency for each dataset. Specifically, for an input sample, we calculate the average similarity among the explanation results generated by  the five approaches. Then, the consistency of a dataset is represented as the average similarity for all samples. The consistency results with $k$ from 1 to 20 are presented in Fig. \ref{Fig-RQ1-Inconsistency}. We can observe that the consistency result increases first and then begins to be stable when $k$ is higher than 5.  
Unfortunately, the consistencies in both the MD datasets and the FI datasets are lower than 0.3, indicating that the explanation results are quite different even for the same classifier model and same input.

\vspace{1em}
\noindent\fbox{
	\parbox{0.95\linewidth}{
		\textbf{Answer to RQ1}:
	 The consistencies for the interpretations with different explanation approaches are lower than 0.3, making analysts hard to select the proper explanation result. The low consistencies present the significant motivation of our work.    
	}
}

\subsection{RQ2: Can the explanation approaches provide stable explanation results?}
\label{subsec_exp_stability}

To answer RQ2, firstly, we need to construct a set of similar models. To this end, we only change the tree numbers when training the RF classifiers while keeping the other arguments (e..g, min\_samples\_leaf=1, and min\_samples\_split=2) as the same. Specifically, for each dataset, we construct five RF classifiers, of which the tree numbers are 98, 99, 100, and 101. Note that, the total loss of the four classifiers are nearly the same, so well as their classification performance. Then, we apply the five explanation approaches on the constructed classifier models and obtain the corresponding explanation results. After that, the average stability for each dataset is calculated according to our proposed stability metric.
Fig. \ref{Fig-RQ2-Stability-MD} and Fig. \ref{Fig-RQ2-Stability-FI} illustrate the stability of the five  approaches on different datasets with $k$ from 1 to 20.  
From the two figures we can observe that: 
\begin{itemize}
	\item {For malware detection, LIME shows the best stability among the five approaches in four datasets (MD-I, MD-III, MD-IV, and MD-V). In MD-II, the stability of LIME varies a lot with different $k$. It increases with the increment of $k$ and starts to be stable around 0.85 when $K$ is higher than 7.}
	\item {For familial identification, the stability of SHAP is similar to that of LIME, which is around 0.85.}
	\item {LEMNA presents the worst stability which is lower than 0.4 on average.} 
\end{itemize}

\begin{figure}[!t]
	\centering
	\includegraphics[width=0.5\textwidth]{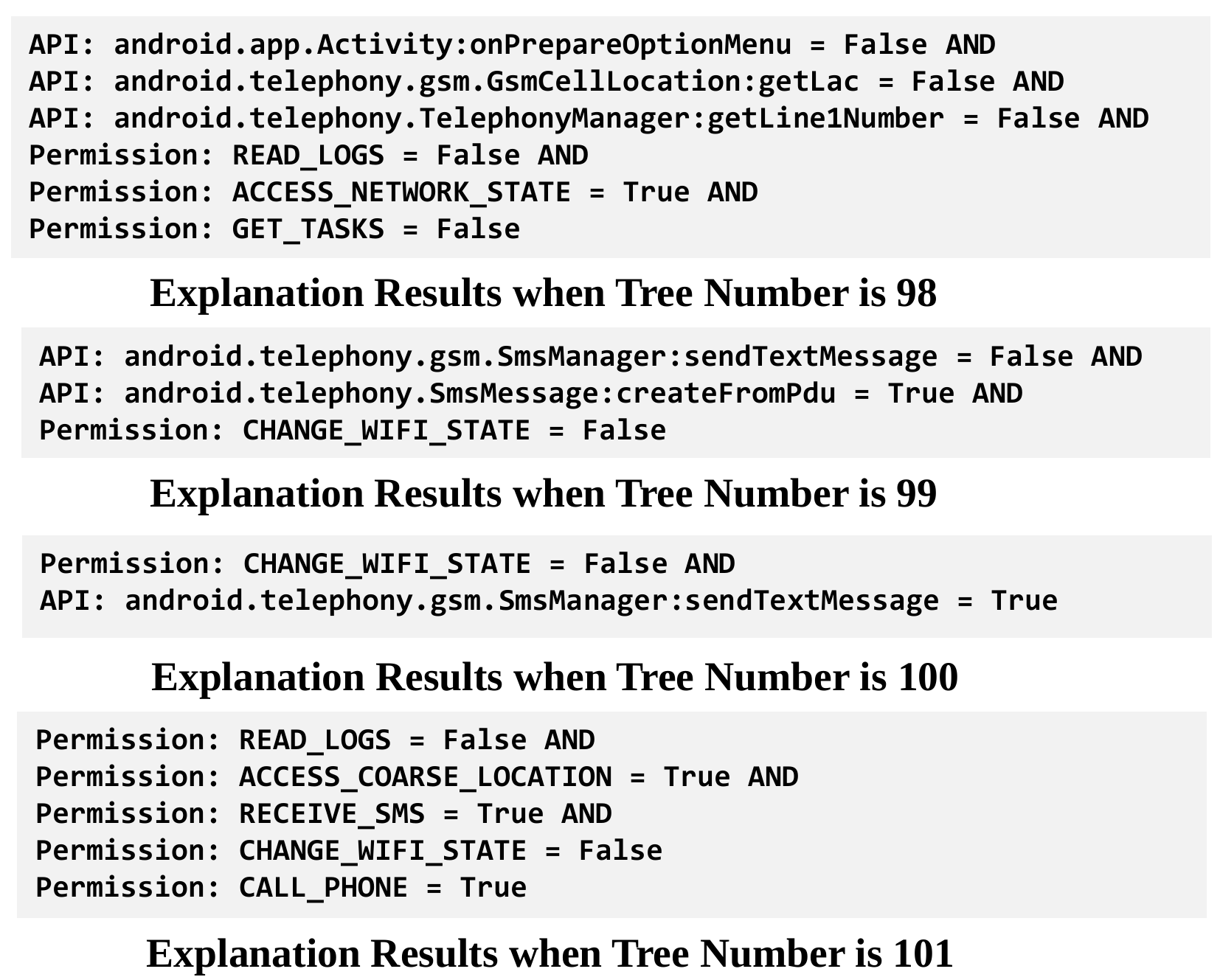}\\
	\caption{Four explanation results for a specific app in FI-I dataset that is classified into the GoldDream family with four similar RF classifier using LORE.}
	\label{Fig-StabilityExample}
	\vspace{-15pt}
\end{figure}

\begin{table*}[!t]
	\centering
	\scriptsize
	\caption{Top 10 features ranked by information gain for four similar RF classifier that are constructed with different tree numbers on the MD-I dataset.}
	\scalebox{0.95}{
		\begin{tabular}{llll} \\ \toprule[1.5pt]
			RF tree number=98 & RF tree number=99 & RF tree number=100 & RF tree number=101 \\ \hline
			READ\_SMS                  & READ\_SMS                  & getSubscriberId             & READ\_SMS                   \\
			getSubscriberId            & getSubscriberId            & READ\_SMS                   & getSubscriberId             \\
			BOOT\_COMPLETED            & BOOT\_COMPLETED            & BOOT\_COMPLETED             & BOOT\_COMPLETED             \\
			getLine1Number             & getLine1Number             & getLine1Number              & SEND\_SMS                   \\
			SEND\_SMS                  & getDeviceId                & SEND\_SMS                   & getLine1Number              \\
			getDeviceId                & SEND\_SMS                  & READ\_PHONE\_STATE          & getDeviceId                 \\
			READ\_PHONE\_STATE         & addView                    & getDeviceId                 & READ\_PHONE\_STATE          \\
			addView                    & CHANGE\_WIFI\_STATE        & RECEIVE\_SMS                & addView                     \\
			RECEIVE\_SMS               & READ\_PHONE\_STATE         & CHANGE\_WIFI\_STATE         & RECEIVE\_SMS                \\
			CHANGE\_WIFI\_STATE        & RECEIVE\_SMS               & addView                     & CHANGE\_WIFI\_STATE  \\ \bottomrule[1.5pt]       
		\end{tabular}
	}
	\label{Tab-featureRanking-DiffTree}
\end{table*}

\begin{table*}[!t]
	\centering
	\scriptsize
	\caption{Top 10 features ranked by information gain for four similar RF classifier that are constructed with fixed 100 tree number on the MD-I dataset four times.}
	\scalebox{0.95}{
		\begin{tabular}{cccc} \\ \toprule[1.5pt]
			RF run=1            & RF run=2            & RF run=3            & RF run=4            \\ \hline
			READ\_SMS           & getSubscriberId     & READ\_SMS           & READ\_SMS           \\
			getSubscriberId     & READ\_SMS           & getSubscriberId     & getSubscriberId     \\
			BOOT\_COMPLETED     & BOOT\_COMPLETED     & BOOT\_COMPLETED     & BOOT\_COMPLETED     \\
			getLine1Number      & getLine1Number      & getLine1Number      & SEND\_SMS           \\
			getDeviceId         & SEND\_SMS           & SEND\_SMS           & getLine1Number      \\
			SEND\_SMS           & getDeviceId         & getDeviceId         & CHANGE\_WIFI\_STATE \\
			CHANGE\_WIFI\_STATE & RECEIVE\_SMS        & RECEIVE\_SMS        & getDeviceId         \\
			RECEIVE\_SMS        & addView             & CHANGE\_WIFI\_STATE & RECEIVE\_SMS        \\
			READ\_PHONE\_STATE  & READ\_PHONE\_STATE  & READ\_PHONE\_STATE  & READ\_PHONE\_STATE  \\
			addView             & CHANGE\_WIFI\_STATE & addView             & addView            \\ \bottomrule[1.5pt] 
		\end{tabular}
	}
	\label{Tab-featureRanking-SameTree}
		\vspace{-10pt}
\end{table*}

Fig. \ref{Fig-StabilityExample} presents an example of the explanation results generated by LORE. Even with similar classifier models, the explanation results of LORE are quite different, i.e., the stability of the samples is only 0.186 when $k=6$. Moreover, for the  second classifier (tree number is 99) and the third classifier (tree number is 100), their explanation results exist a conflict, i.e., the value of API \texttt{sendTextMessage} is false in the former while true in the latter.

\begin{figure}[!t]
		\centering
		\includegraphics[scale=0.4]{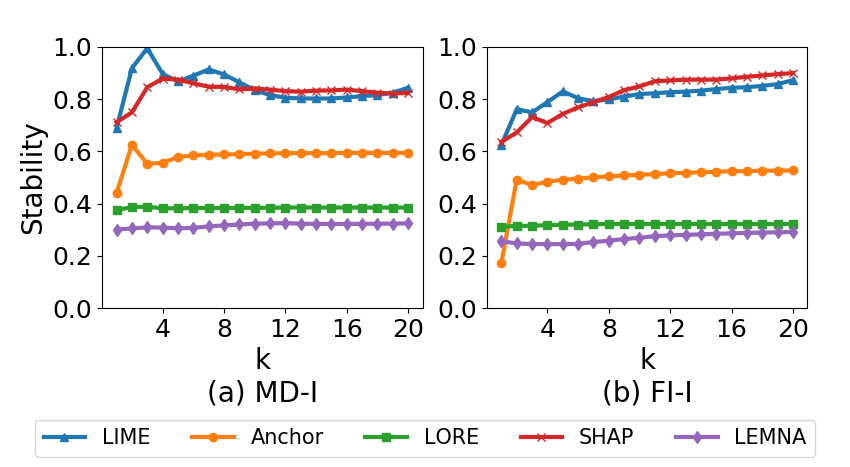}
	\caption{Stability of the five explanation approaches on MD-I and FI-I datasets with $k$ from 1 to 20 when training RF classifiers with fixed 100 trees.}
	\label{Fig-RQ2-Stability-SameTree}
	\vspace{-18pt}
\end{figure}

Moreover, we also assess the stabilities on four RF classifiers trained only with different random seeds when the  tree number is fixed as 100. The results are presented in Fig. \ref{Fig-RQ2-Stability-SameTree}, which demonstrate that when the tree numbers are fixed as 100, the stabilities of the explanation approaches are similar to those evaluated on the similar models trained with different tree numbers. LIME and SHAP could also achieve higher stabilities than the other three approaches.

We also calculate the top 10 features of similar RF classifiers ranked by information gain. The results are listed in Table \ref{Tab-featureRanking-DiffTree} and Table \ref{Tab-featureRanking-SameTree}. Note that in Table \ref{Tab-featureRanking-DiffTree}, the four similar classifiers are constructed with different tree numbers on the same training dataset. In Table \ref{Tab-featureRanking-SameTree}, the four similar classifiers are constructed with a fixed 100 tree number on the same training dataset four times. We can observe that for the classifiers in both Table \ref{Tab-featureRanking-DiffTree} and Table \ref{Tab-featureRanking-SameTree}, their top ten features are nearly the same. The results demonstrate that these classifiers present very similar behaviors.

However, the stabilities of Anchor, LORE, and LEMNA are lower than 0.6, indicating that their explanation results vary a lot on similar models. Take LEMNA as an example, when k is selected as 10, there would be less than 3 common explanation features between multiple runs. In addition, the stabilities of LIME and SHAP are generally higher than 0.8, which makes them more reliable than other approaches. We think that the different stabilities are mainly caused by the explanation approaches themselves rather than the used learning algorithms.


\vspace{1em}
\noindent\fbox{
	\parbox{0.95\linewidth}{
		\textbf{Answer to RQ2}: The ranking of the five explaining approaches in term of the stability metric is  LIME $\ge$ SHAP $>$ Anchor $>$ LORE $>$ LEMNA. The stabilities of LIME and SHAP are around 0.85, indicating that their explanation features hardly change when slightly modify the model.
	}
}

\begin{figure*}[!t]
	\centering
	\includegraphics[width=\textwidth]{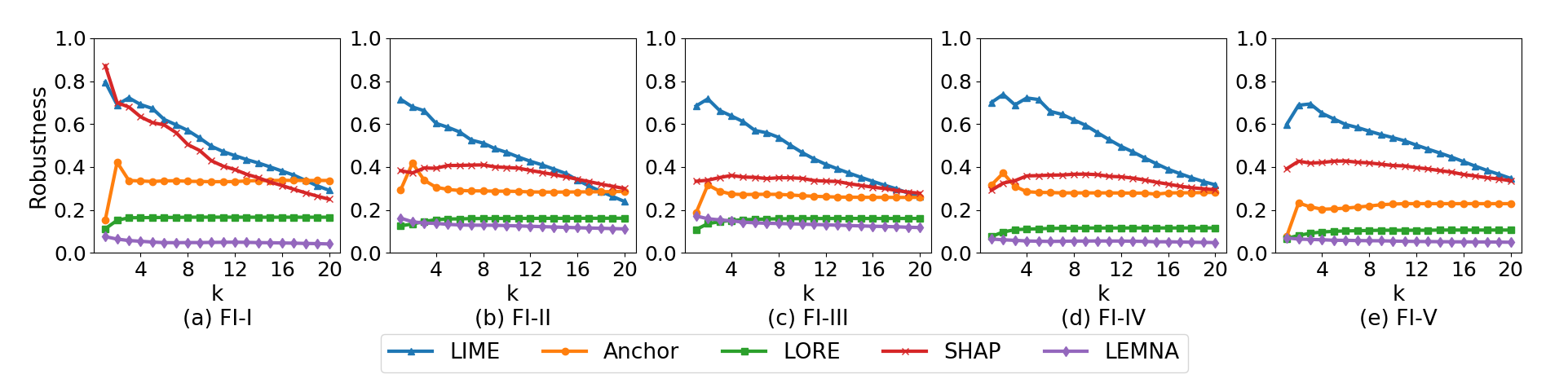}\\
	\caption{Robustness of the five explaining methods on different familial identification datasets with $k$ from 1 to 20.}
	\label{Fig-RQ3-Robustness-FI}
		\vspace{-10pt}
\end{figure*}

\begin{table*}[!t]
	
	\centering
	\scriptsize
	\caption{Robustness score of the interpretations generated by five explanation approaches for different families in FI-III testing dataset when $k=5$.}
	\scalebox{1.0}{
		\begin{tabular}{cc|ccc|ccc|ccc|ccc|ccc} \\ \toprule[1.5pt]
			\multicolumn{1}{c}{\multirow{2}{*}{Malware Family}} & \multicolumn{1}{c}{\multirow{2}{*}{\#Sample}} & \multicolumn{3}{|c}{LIME}                                                       & \multicolumn{3}{|c}{Anchor}                 & \multicolumn{3}{|c}{LORE}   & \multicolumn{3}{|c}{SHAP}  & \multicolumn{3}{|c}{LEMNA} \\
			\multicolumn{1}{c}{}                                & \multicolumn{1}{c}{}                               & \multicolumn{1}{|c}{$\mathcal{\overline{S}}$} & \multicolumn{1}{c}{$\mathcal{\overline{D}}$} & \multicolumn{1}{c}{$rob$} & \multicolumn{1}{|c}{$\mathcal{\overline{S}}$} & $\mathcal{\overline{D}}$     & $rob$ & 
			$\mathcal{\overline{S}}$     & $\mathcal{\overline{D}}$     & $rob$ &
			$\mathcal{\overline{S}}$     & $\mathcal{\overline{D}}$     & $rob$ &
			$\mathcal{\overline{S}}$     & $\mathcal{\overline{D}}$     & $rob$\\ \hline
			adwo                                                & 169                                                & 0.704                 & 0.222                 & 0.482                          & 0.272                 & 0.077 & 0.195      & 0.2   & 0.099 & 0.101      & 0.281 & 0.117 & 0.164    &  0.079  &   0.034   &  0.045  \\
			airpush                                             & 38                                                 & 0.837                 & 0.234                 & 0.603                          & 0.218                 & 0.079 & 0.139      & 0.251 & 0.132 & 0.119      & 0.336 & 0.171 & 0.165   &  0.146   &  0.05    &    0.096   \\
			anserver                                            & 27                                                 & 0.911                 & 0.244                 & 0.667                          & 0.363                 & 0.051 & 0.312      & 0.171 & 0.095 & 0.076      & 0.527 & 0.089 & 0.438    &   0.111 &   0.041  &   0.07     \\
			basebridge                                          & 152                                                & 0.892                 & 0.239                 & 0.653                          & 0.311                 & 0.054 & 0.257      & 0.163 & 0.088 & 0.075      & 0.526 & 0.136 & 0.39   &  0.094  &   0.042   &  0.052    \\
			boqx                                                & 25                                                 & 0.689                 & 0.193                 & 0.496                          & 0.183                 & 0.056 & 0.127      & 0.207 & 0.092 & 0.115      & 0.309 & 0.098 & 0.211    &    0.138   &    0.044 &   0.094  \\
			boxer                                               & 48                                                 & 0.865                 & 0.27                  & 0.595                          & 0.301                 & 0.062 & 0.239      & 0.238 & 0.089 & 0.149      & 0.648 & 0.144 & 0.504   &  0.187   &   0.053   &    0.134     \\
			clicker                                             & 19                                                 & 0.616                 & 0.168                 & 0.448                          & 0.244                 & 0.079 & 0.165      & 0.228 & 0.09  & 0.138      & 0.323 & 0.116 & 0.207   &  0.050   &  0.040    &   0.010    \\
			dowgin                                              & 425                                                & 0.812                 & 0.061                 & 0.751                          & 0.327                 & 0.043 & 0.284      & 0.245 & 0.069 & 0.176      & 0.504 & 0.067 & 0.437    &   0.067   &    0.029    &    0.038  \\
			droiddreamlight                                     & 51                                                 & 0.705                 & 0.293                 & 0.412                          & 0.177                 & 0.088 & 0.089      & 0.417 & 0.138 & 0.279      & 0.578 & 0.141 & 0.437   &   0.193  &   0.029   &   0.164    \\
			droidkungfu                                         & 368                                                & 0.815                 & 0.256                 & 0.559                          & 0.349                 & 0.092 & 0.257      & 0.171 & 0.106 & 0.065      & 0.404 & 0.102 & 0.302    &   0.104  &   0.039   &   0.065  \\
			droidsheep                                          & 7                                                  & 0.781                 & 0.108                 & 0.673                          & 0.057                 & 0.047 & 0.01       & 0.285 & 0.112 & 0.173      & 0.885 & 0.063 & 0.822    &   0.238  &   0.032  &   0.206   \\
			fakeangry                                           & 8                                                  & 0.686                 & 0.101                 & 0.585                          & 0.171                 & 0.061 & 0.11       & 0.158 & 0.117 & 0.041      & 0.8   & 0.097 & 0.703   &   0.060  &  0.030  &  0.030     \\
			fakedoc                                             & 74                                                 & 0.641                 & 0.023                 & 0.618                          & 0.235                 & 0.036 & 0.199      & 0.155 & 0.088 & 0.067      & 0.729 & 0.068 & 0.661   &  0.032   &   0.032  & 0.000   \\
			fakeinst                                            & 752                                                & 0.94                  & 0.287                 & 0.653                          & 0.427                 & 0.074 & 0.353      & 0.242 & 0.068 & 0.174      & 0.486 & 0.069 & 0.417  &  0.195  &  0.041  &  0.154     \\
			fakeplay                                            & 22                                                 & 0.791                 & 0.122                 & 0.669                          & 0.199                 & 0.088 & 0.111      & 0.306 & 0.067 & 0.239      & 0.8   & 0.097 & 0.703   &  0.347   &  0.032   &  0.315   \\
			geinimi                                             & 53                                                 & 0.749                 & 0.081                 & 0.668                          & 0.137                 & 0.047 & 0.09       & 0.142 & 0.076 & 0.066      & 0.736 & 0.076 & 0.66  &  0.042  & 0.033   & 0.009     \\
			gingermaster                                        & 193                                                & 0.817                 & 0.316                 & 0.501                          & 0.401                 & 0.107 & 0.294      & 0.277 & 0.096 & 0.181      & 0.407 & 0.128 & 0.279  &  0.156  &  0.031  &  0.125    \\
			golddream                                           & 40                                                 & 0.768                 & 0.205                 & 0.563                          & 0.113                 & 0.045 & 0.068      & 0.142 & 0.096 & 0.046      & 0.547 & 0.124 & 0.423  & 0.048  &  0.031  &  0.017    \\
			hongtoutou                                          & 23                                                 & 0.762                 & 0.234                 & 0.528                          & 0.192                 & 0.086 & 0.106      & 0.269 & 0.099 & 0.17       & 0.528 & 0.187 & 0.341   &  0.242  &  0.038  & 0.204   \\
			iconosys                                            & 77                                                 & 0.956                 & 0.371                 & 0.585                          & 0.425                 & 0.117 & 0.308      & 0.261 & 0.101 & 0.16       & 0.713 & 0.151 & 0.562    & 0.107  &  0.034   & 0.073  \\
			imlog                                               & 21                                                 & 0.872                 & 0.182                 & 0.69                           & 0.135                 & 0.053 & 0.082      & 0.474 & 0.132 & 0.342      & 0.892 & 0.147 & 0.745    & 0.139  &  0.042  &  0.097    \\
			jsmshider                                           & 11                                                 & 0.769                 & 0.142                 & 0.627                          & 0.105                 & 0.05  & 0.055      & 0.265 & 0.057 & 0.208      & 0.825 & 0.109 & 0.716    &  0.389  &   0.048  &  0.341  \\
			kmin                                                & 124                                                & 0.754                 & 0.232                 & 0.522                          & 0.204                 & 0.056 & 0.148      & 0.217 & 0.061 & 0.156      & 0.597 & 0.105 & 0.492  &  0.056  &   0.031  &   0.025     \\
			kuguo                                               & 179                                                & 0.842                 & 0.227                 & 0.615                          & 0.377                 & 0.074 & 0.303      & 0.285 & 0.086 & 0.199      & 0.282 & 0.101 & 0.181   &  0.564   &  0.041  & 0.523   \\
			lovetrap                                            & 10                                                 & 0.787                 & 0.104                 & 0.683                          & 0.088                 & 0.049 & 0.039      & 0.157 & 0.122 & 0.035      & 0.68  & 0.157 & 0.523     &   0.062  &  0.034  &   0.028  \\
			mobiletx                                            & 41                                                 & 0.851                 & 0.214                 & 0.637                          & 0.175                 & 0.071 & 0.104      & 0.226 & 0.088 & 0.138      & 1     & 0.125 & 0.875 &  0.098  &  0.024  &  0.074     \\
			pjapps                                              & 41                                                 & 0.748                 & 0.356                 & 0.392                          & 0.281                 & 0.078 & 0.203      & 0.222 & 0.105 & 0.117      & 0.698 & 0.137 & 0.561 &  0.119  &  0.029  &   0.090     \\
			plankton                                            & 448                                                & 0.874                 & 0.285                 & 0.589                          & 0.415                 & 0.067 & 0.348      & 0.331 & 0.091 & 0.24       & 0.214 & 0.109 & 0.105   &  0.393  &  0.028  &   0.365   \\
			smskey                                              & 56                                                 & 0.532                 & 0.223                 & 0.309                          & 0.259                 & 0.079 & 0.18       & 0.259 & 0.081 & 0.178      & 0.477 & 0.09  & 0.387   &  0.157  &  0.047   &  0.110     \\
			smsreg                                              & 75                                                 & 0.758                 & 0.255                 & 0.503                          & 0.184                 & 0.068 & 0.116      & 0.167 & 0.102 & 0.065      & 0.307 & 0.105 & 0.202    &   0.035    &   0.038  &  -0.003  \\
			steek                                               & 10                                                 & 0.831                 & 0.011                 & 0.82                           & 0.08                  & 0.032 & 0.048      & 0.346 & 0.111 & 0.235      & 0.8   & 0.093 & 0.707     &  0.040  &   0.017  &  0.023 \\
			utchi                                               & 143                                                & 0.977                 & 0.133                 & 0.844                          & 0.479                 & 0.049 & 0.43       & 0.388 & 0.107 & 0.281      & 0.921 & 0.109 & 0.812     &   0.196   &  0.038   & 0.158 \\
			waps                                                & 386                                                & 0.875                 & 0.251                 & 0.624                          & 0.345                 & 0.066 & 0.279      & 0.184 & 0.101 & 0.083      & 0.272 & 0.093 & 0.179    &  0.170   &   0.034   &  0.136  \\
			youmi                                               & 57                                                 & 0.768                 & 0.226                 & 0.542                          & 0.274                 & 0.079 & 0.195      & 0.254 & 0.112 & 0.142      & 0.23  & 0.133 & 0.097 &   0.232  &   0.031   &  0.201     \\
			yzhc                                                & 25                                                 & 0.627                 & 0.166                 & 0.461                          & 0.13                  & 0.048 & 0.082      & 0.327 & 0.096 & 0.231      & 0.753 & 0.118 & 0.635    &  0.117   &  0.054  &  0.063  \\
			zitmo                                               & 15                                                 & 0.701                 & 0.083                 & 0.618                          & 0.093                 & 0.032 & 0.061      & 0.222 & 0.073 & 0.149      & 0.516 & 0.085 & 0.431 &   0.164  &   0.046 &  0.118  \\ 
			\textbf{Average}                                             &                                                &     \textbf{0.840}                      &     \textbf{0.229}                      & \textbf{0.611}                          &  \textbf{0.340}                     &   \textbf{0.069}    & \textbf{0.271}      &   \textbf{0.242}    &  \textbf{0.089}     & \textbf{0.154}      &   \textbf{0.452}    &  \textbf{0.099}     & \textbf{0.354}  &  
			\textbf{0.179}    &    \textbf{0.035}  &  \textbf{0.144}\\ \bottomrule[1.5pt]
		\end{tabular}
	}
	\label{Tab-familyRobust}
		\vspace{-15pt}
\end{table*}

\subsection{RQ3: Can the explanation approaches provide robust interpretations?}
\label{subsec_exp_robustness}

As introduced before, high robustness indicates that the samples with the same predicted labels contain similar interpretations, while the samples with different predicted labels contain different interpretations. For example, in familial identification, the sample in family \textit{geinimi} should present similar interpretations with the samples within the same family. 

To answer RQ3, we only use the interpretations generated with the RF classifier model, of which the tree number is 100. After that, we calculate the robustness score for each explaining method on different datasets. The results are illustrated in Fig. \ref{Fig-RQ3-Robustness-FI}.  From the figure we can observe that:

\begin{itemize}
	\item {LIME shows the best robustness among the five approaches on all the datasets. However, with an increase of $k$, the robustness of LIME decreases.}
	\item {The robustness of SHAP differs from the datasets. In FI-I, it presents similar robustness with LIME. However, in the other three FI datasets, its robustness scores are around 0.4.}
	\item {For Anchor, LORE, and LEMNA, their robustness score are lower than 0.4, and they hardly change with the $k$ value.}
\end{itemize}

We further investigate the robustness scores in different families. Due to the page limitation, we only present the robustness scores for the 36 families in FI-III dataset when $k$ is set as 5. As listed in Table \ref{Tab-familyRobust},  the first column and the second column list the family name and the size of corresponding test samples, respectively. Columns 3-14 list the $\mathcal{\overline{S}}$, $\mathcal{\overline{D}}$, and $rob$ for each family with different explaining methods. Here, $\mathcal{\overline{S}}$ and $\mathcal{\overline{D}}$ denote the average value of the $avg(x_i,\mathcal(S(x_i)))$ and  $avg(x_i,\mathcal(D(x_i)))$ for all samples with a specific family, respectively. In addition, $rob=\mathcal{\overline{S}}-\mathcal{\overline{D}}$. From Table \ref{Tab-familyRobust}, we can observe that:
\begin{itemize}
	\item {For LIME, the highest $\mathcal{\overline{S}}$ is 0.977 for the \textit{utchi} family, and the lowest $\mathcal{\overline{S}}$ can also achieve 0.532, which is higher than all the $\mathcal{\overline{S}}$ in Anchor and LORE. The average value of $\mathcal{\overline{S}}$ is 0.840, indicating that when $k=5$, more than 4 explanation features are same for the samples within the same family.}
	\item{For the $\mathcal{\overline{D}}$ metric, most  $\mathcal{\overline{D}}$ for Anchor, LORE, and LEMNA is lower than 0.1, indicating that the explanations generated by such three approaches are quite different.
	Even the $avg(x_i,\mathcal(D(x_i)))$ for LIME is around 0.229, we think it is an acceptable value since most malware samples would have common malicious behaviors.}
	\item{Although the average $rob$ value of SHAP is only 0.354, it differs a lot with families. For example, the $rob$ can achieve 0.875 in \textit{mobiletx} family while is only 0.097 in \textit{youmi} family.}
	\item{For LEMNA, $\mathcal{\overline{S}}$ and $\mathcal{\overline{D}}$ of some families (e.g., \textit{fakedoc} and \textit{smsreg}) are nearly the same, indicating that this approach cannot capture the core differences between malware families.}
\end{itemize}


\begin{figure}[!t]
	\centering
	\includegraphics[width=0.5\textwidth]{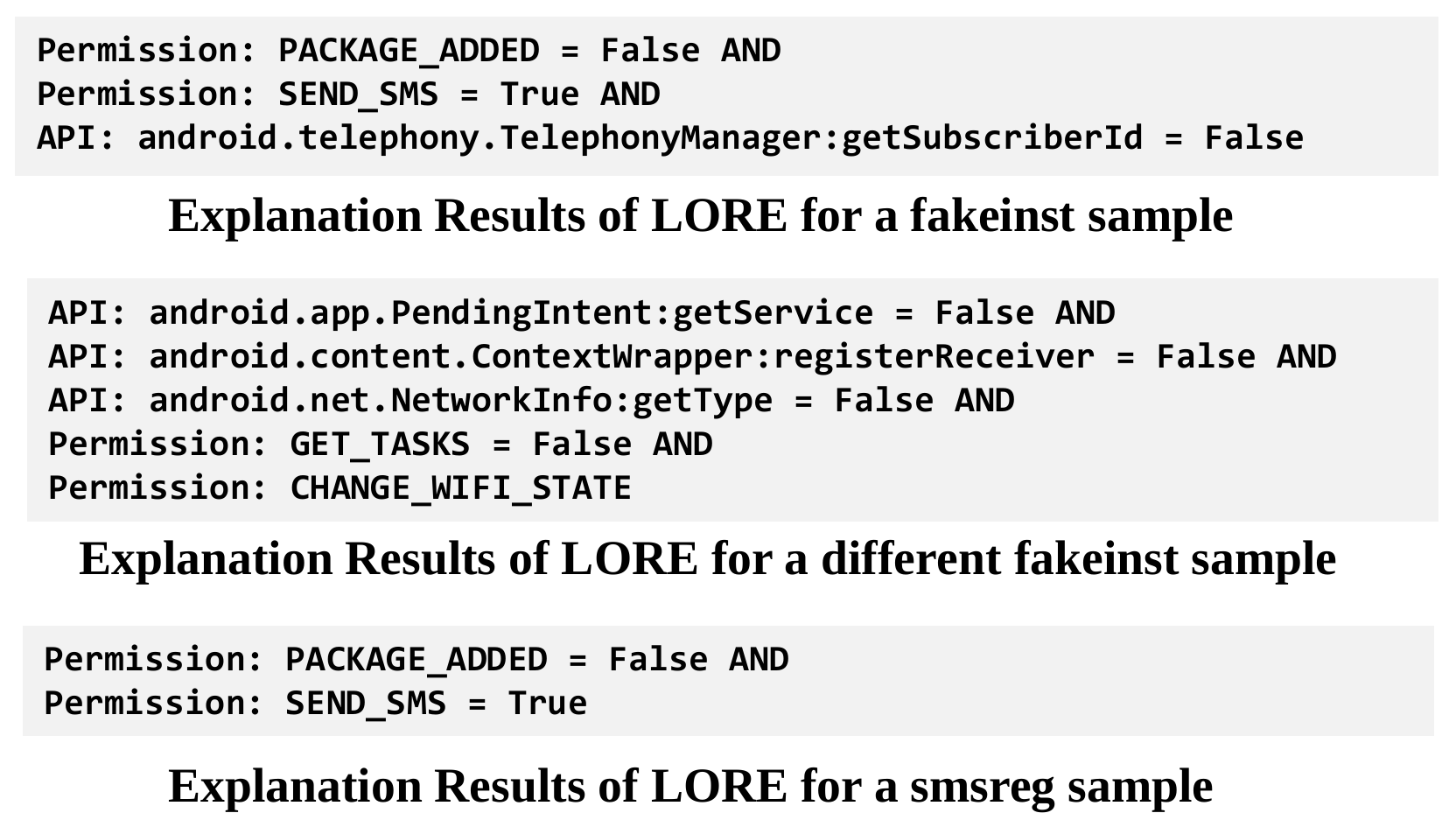}\\
	\caption{Explanation results for two \textit{fakeinst} samples and one \textit{smsreg} sample in FI-III datast using LORE.}
	\label{Fig-RobustnessExample}
	\vspace{-20pt}
\end{figure}
\begin{figure*}[!t]
	\centering
	\includegraphics[width=\textwidth]{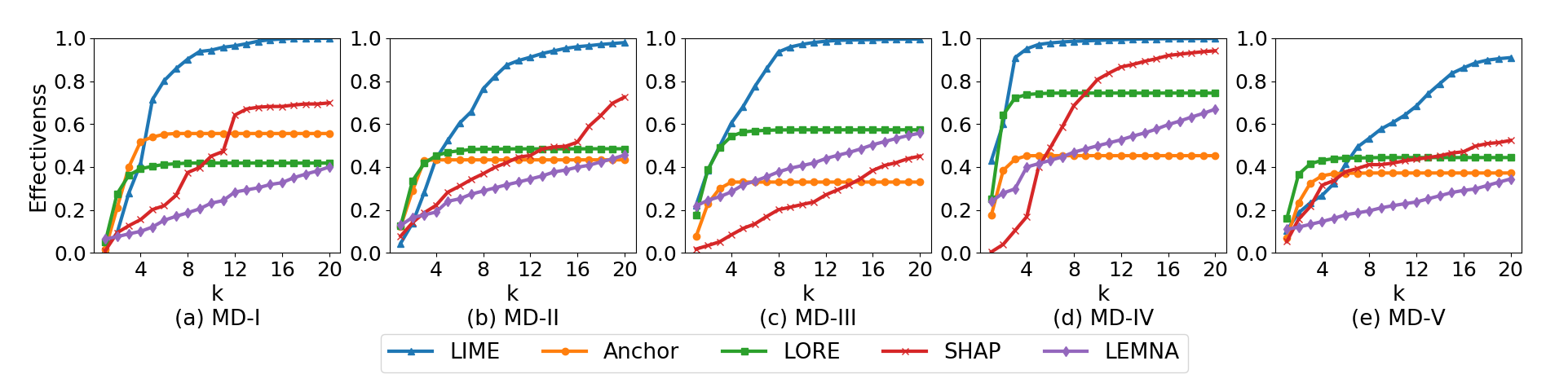}\\
	\caption{Effectiveness of the five explaining methods on different malware detection datasets with $k$ from 1 to 20.}
	\label{Fig-RQ4-Effectiveness-MD}
	\vspace{-12pt}
\end{figure*}

\begin{figure*}[!t]
	\centering
	\includegraphics[width=\textwidth]{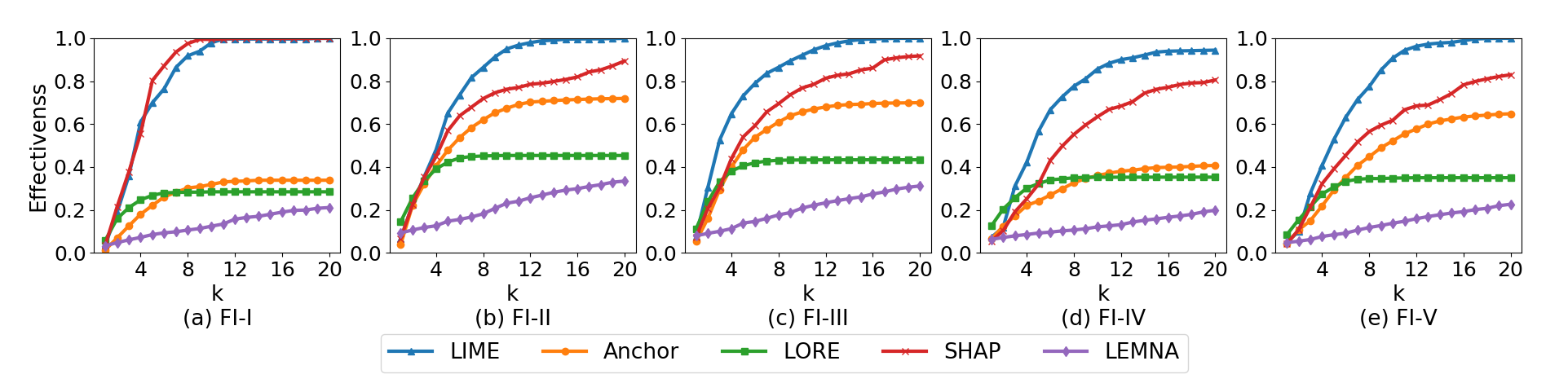}\\
	\caption{Effectiveness of the five explaining methods on different familial identification datasets with $k$ from 1 to 20.}
	\label{Fig-RQ4-Effectiveness-FI}
	\vspace{-8pt}
\end{figure*}

Based on the above results, we can identify that nearly all the approaches are fragile to be attacked by small perturbations to the input sample while not changing the label. Here we use three samples to show the frailness of LORE, in which two samples present similar malicious behaviors and are classified into the same \textit{fakeinst} family. The difference between the input feature vectors is regarded as the small perturbations. The third sample is classified into the \textit{smsreg} family. The explanation results generated by LORE are presented in Fig. \ref{Fig-RobustnessExample}. The results demonstrate that for the two samples in the same family, their explanation similarity is 0. However, for the first and the third samples, even their labels are different, their explanation similarity can achieve 0.8.

\vspace{1em}
\noindent\fbox{
	\parbox{0.95\linewidth}{
		\textbf{Answer to RQ3}: The ranking of the five explaining approaches in term of the robustness metric is LIME $>$ SHAP $>$ Anchor $>$ LORE $>$ LEMNA. Moreover, the robustness scores of these explanation approaches are not high enough, indicating that their interpretations for samples with same labels are easy to be altered.   
	}
}



\begin{table*}[!t]
	\centering
	\scriptsize
	\caption{Ranking of the effective explanation features generated by five explanation approaches on different malware detection datasets when $k=5$.}
	\scalebox{0.8}{
		\begin{tabular}{lll|ll|ll|ll|ll}  \\ \toprule[1.5pt]
			& \multicolumn{2}{c|}{LIME}           & \multicolumn{2}{c|}{Anchor}         & \multicolumn{2}{c|}{LORE}           & \multicolumn{2}{c}{SHAP}   & \multicolumn{2}{c}{LEMNA}        \\ 
			& Feature                  & Weight & Feature                  & Weight & Feature                  & Weight & Feature                  & Weight & Feature                  & Weight\\ \hline
			\multirow{5}{*}{MD-I}                      & getSubscriberId          & 0.71    & READ\_SMS                & 0.257   & getLine1Number           & 0.141   & READ\_SMS                & 0.174   & READ\_SMS   &   0.032\\
			& BOOT\_COMPLETED          & 0.677   & getSubscriberId          & 0.249   & CHANGE\_WIFI\_STATE      & 0.101   & getSubscriberId          & 0.153  & getSubscriberId   &  0.027 \\
			& READ\_SMS                & 0.548   & BOOT\_COMPLETED          & 0.179   & READ\_SMS                & 0.086   & READ\_SMS                & 0.056   & getDeviceId  &  0.025\\
			& getDeviceId              & 0.406   & READ\_PHONE\_STATE       & 0.112   & getSubscriberId          & 0.085   & getLac                   & 0.035    &  READ\_LOGS   &   0.018\\
			& addView                  & 0.358   & SEND\_SMS                & 0.092   & getDeviceId              & 0.071   & getLine1Number           & 0.029   & getLine1Number   &  0.017 \\ \hline
			\multirow{5}{*}{MD-II}                     & SEND\_SMS                & 0.521   & READ\_PHONE\_STATE       & 0.181   & getDeviceId              & 0.194   & getDeviceId              & 0.172  &   READ\_PHONE\_STATE    &  0.122  \\
			& READ\_PHONE\_STATE       & 0.463   & SEND\_SMS                & 0.139   & READ\_PHONE\_STATE       & 0.185   & SEND\_SMS                & 0.156  & SEND\_SMS  &   0.074 \\
			& addView                  & 0.349   & getLine1Number           & 0.091   & addView                  & 0.094   & READ\_PHONE\_STATE       & 0.082   & WRITE\_CONTACTS & 0.042\\
			& READ\_SMS                & 0.195   & getSubscriberId          & 0.086   & SEND\_SMS                & 0.092   & getLine1Number           & 0.077  & ACCESS\_NETWORK\_STATE  &  0.040 \\
			& getDeviceId              & 0.139   & addView                  & 0.083   & getType                  & 0.067   & getSubscriberId          & 0.062   & ACCESS\_WIFI\_STATE    &   0.039\\ \hline
			\multirow{5}{*}{MD-III}                    & READ\_PHONE\_STATE       & 0.681   & SEND\_SMS                & 0.106   & READ\_PHONE\_STATE       & 0.388   & SEND\_SMS                & 0.077   &    READ\_PHONE\_STATE   &   0.275 \\
			& getSubscriberId          & 0.445   & READ\_PHONE\_STATE       & 0.079   & getType                  & 0.185   & getSubscriberId          & 0.058  &  SEND\_SMS   &   0.053  \\
			& getDeviceId              & 0.277   & RECEIVE\_SMS             & 0.069   & getDeviceId              & 0.114   & READ\_PHONE\_STATE       & 0.044  &  READ\_EXTERNAL\_STORSGE  & 0.042 \\
			& SEND\_SMS                & 0.22    & addView                  & 0.061   & SEND\_SMS                & 0.064   & getDeviceId              & 0.044  & WRITE\_CONTACTS & 0.031 \\
			& getType                  & 0.056   & getDeviceId              & 0.055   & GET\_ACCOUNTS            & 0.051   & sendTextMessage          & 0.025  & ACCESS\_WIFI\_STATE & 0.024 \\ \hline
			\multicolumn{1}{c}{\multirow{5}{*}{MD-IV}} & READ\_PHONE\_STATE       & 0.971   & READ\_PHONE\_STATE       & 0.299   & READ\_PHONE\_STATE       & 0.571   & getDeviceId              & 0.364  &   READ\_PHONE\_STATE  & 0.386  \\
			\multicolumn{1}{c}{}                       & COARSE\_LOCATION & 0.48    & COARSE\_LOCATION & 0.143   & getDeviceId              & 0.22    & getMacAddress            & 0.313  &   WRITE\_SETTINGS  &  0.115   \\
			\multicolumn{1}{c}{}                       & getDeviceId              & 0.428   & getLine1Number           & 0.064   & getMacAddress            & 0.103   & COARSE\_LOCATION & 0.271   &   BOOT\_COMPLETED  &  0.087 \\
			\multicolumn{1}{c}{}                       & getMacAddress            & 0.057   & SEND\_SMS                & 0.048   & ACCESS\_WIFI\_STATE      & 0.088   & getLine1Number           & 0.091  &   READ\_EXTERNAL\_STORSGE  &   0.035\\
			\multicolumn{1}{c}{}                       & ACCESS\_WIFI\_STATE      & 0.023   & getSimOperatorName       & 0.037   & COARSE\_LOCATION & 0.073   & SEND\_SMS                & 0.089   &   getDeviceId  &   0.015\\  \hline
			\multirow{5}{*}{MD-V}   &   READ\_PHONE\_STATE      &    0.323    &   getLine1Number      &    0.095          &  READ\_PHONE\_STATE     &     0.264          &   SEND\_SMS    &   0.192       &   READ\_PHONE\_STATE    &   0.079       \\  
			                        &   SEND\_SMS   &    0.202   &        getSubscriberId        &        0.085         &    getDeviceId     &   0.073           &   getLine1Number      &  0.154       &    SEND\_SMS   &  0.029    \\
			                        &    getLine1Number  &  0.141  &     SEND\_SMS        &        0.076        &     getLine1Number      &   0.071         &    BOOT\_COMPLETED    & 0.151             &  BOOT\_COMPLETED  &   0.023  \\
			                        &    getSubscriberId  & 0.096  &    BOOT\_COMPLETED        &   0.069             &    getSubscriberId       &   0.067          &    RECEIVE\_SMS      &     0.093       &   WRITE\_SETTINGS   &  0.017 \\
			                        &    getDeviceId    &  0.046  &    CHANGE\_WIFI\_STATE       &   0.037               &     SEND\_SMS         &   0.044         &   getSubscriberId       &     0.049      &   getDeviceId  &   0.012  \\ \bottomrule[1.5pt]

		\end{tabular}
	}
	\label{Tab-MDEffect}
\end{table*}

\begin{table}[!t]
	\centering
	\scriptsize
	\caption{Malicious behaviors identified by humans and their corresponding explanation features generated by LIME.}
	\scalebox{0.9}{
		\begin{tabular}{ll} \\ \toprule[1.5pt]
			\multicolumn{1}{c}{Malicious Behaviors}                                                                                                                                                                                                                                                                                                                                                   & Explanation Features                                                                                                             \\ \hline
			\begin{tabular}[c]{@{}l@{}}The \textit{fakeinst} samples generally appear to be \\ installers for various benign apps. While active, \\ the malware samples send SMS messages to\\ premium rate phone numbers.\end{tabular}                                                                                                        & \begin{tabular}[c]{@{}l@{}}getOrininatingAddress;\\ createFromPdu;\\ RECEIVE\_SMS;\\ SEND\_SMS;\\ INSTALL\_PACKAGES\end{tabular} \\ \hline
			\begin{tabular}[c]{@{}l@{}}The \textit{svpeng} samples belong to the type of banking \\ trojan and ransomware. If its C\&C server sends a \\ specific command, it will lock the infected device \\ by using the SYSTEM\_ALERT\_WINDOW permission.\end{tabular}                                                                                                           & \begin{tabular}[c]{@{}l@{}}SYSTEM\_ALERT\_WINDOW;\\ READ\_CONTACTS;\\ getRunningTasks\end{tabular}                              \\ \hline
			\begin{tabular}[c]{@{}l@{}} \textit{slembunk} samples schedule a \textit{java.lang.Runnable} \\ every 4 seconds to monitor the current running activity. \\ Once the activity belongs to a certain banking app, it will \\ overlay a phishing window. Moreover, these samples \\ also forward phone calls and SMS from bank numbers.\end{tabular} & \begin{tabular}[c]{@{}l@{}}RECEIVE\_SMS;\\ CALL\_PHONE;\\ onStartCommand;\\ getMessageBody;\\ SYSTEM\_ALERT\_WINDOW\end{tabular} \\ \bottomrule[1.5pt]
			
		\end{tabular}
	}
	\label{Tab-GrounTruth}
\end{table}

\subsection{RQ4: Can the explanation approaches provide effective interpretations?}
\label{subsec_exp_effectiveness}

The interpretation effectiveness is the most important property of an explanation approach when applying in practice. In our work, the effectiveness is measured by whether the predicted label is changed after the  mutation of  explanation features. 
To answer RQ4, we first mutate the feature values of the given feature vector. Then we feed the new feature vector to the original classifier model and obtain its new predicted label. Finally, we compare the original predicted label and the new predicted label to check the effectiveness of the  explanation features. We evaluate the effectiveness of the five approaches on different datasets. The results are illustrated in Fig. \ref{Fig-RQ4-Effectiveness-MD} and Fig. \ref{Fig-RQ4-Effectiveness-FI}, from which  we can draw three conclusions:

\begin{itemize}
	\item {On the whole, for both the two tasks, LIME shows the best effectiveness among the five approaches. After the mutation for the explanation results generated by LIME, nearly all the predicted labels are changed.}
	\item{With the increase of $k$, the effectiveness of LIME, SHAP, and LEMNA increase. However, for Anchor and LORE, their effectiveness start to be stable when $k>=5$. The main reason is that most of explanation feature size of Anchor and LORE are less than 5, resulting the no change of predicted label with higher $k$ values.}
	\item {The effectiveness of the same approach differs with datasets. For example, LORE performs better than Anchor, SHAP, and LEMNA in MD-II, MD-III, MD-IV and MD-V. However, it performs worse than  Anchor and SHAP in all the four FI datasets.}
\end{itemize}

We further investigate the  generated effective explanation features, which denote the explanation features that can change the predicted label after the mutation of feature values. For example, when $k=5$, the number of explanation feature is 5. If the predicted label is changed after the mutation of the first explanation feature, then the effective explanation feature set contains only one feature. If the predicted label is not changed after the mutation of all the explanation features, then the effective explanation feature set is an empty set. 

Table \ref{Tab-MDEffect} lists the ranking of the effective explanation features generated by five approaches on different MD datasets when $k=5$. Note that for the  API features in this table, we only use their function name due to the page limit. The weight value in the table denotes the significance of each feature to the malware detection. It is obtained based on the percent of malware samples in each dataset that contain the corresponding effective explanation feature. For example, in MD-I dataset, by applying LIME, the weight of API \texttt{getSubscriberId} is 0.71, indicating that  the feature occurs in the effective explanation feature set of 71\% samples in the MD-I testing dataset. 
It is worth noting that in MD-I dataset, the effectiveness of LIME is also 0.71 when $k=5$ according to Fig. \ref{Fig-RQ4-Effectiveness-MD}(a), indicating that for the malware samples in MD-I testing dataset, the API \texttt{getSubscriberId} is the necessary feature to be mutated in order to change the predicted label.

From the table we find that even the listed top-5 features for different explaining methods are similar in the same dataset, their weights are quite different. For example, in MD-IV dataset, the permission \texttt{READ\_PHONE\_STATE} occurs first for LIME, Anchor, LORE and LEMNA. Its weight in LIME is 0.971, which is much higher than those in the other three approaches. The result demonstrates that all the three explanation approaches agree that \texttt{READ\_PHONE\_STATE} is the most important feature for malware detection in this dataset. However, the ability of LIME to capture the true malicious behaviors is better than others.


The above results demonstrate that LIME performs best among the five explanation approaches in terms of the effectiveness metric. We further conduct a comparison between the top-ranked features of LIME and some malware families, e.g., \textit{fakeinst}, \textit{svpeng}, and \textit{slembunk}, to check whether the generated explanation features can depict the malicious behaviors identified by the authors of dataset-V (i.e., AMD dataset~\cite{wei2017deep}). 
The malicious behaviors identified by humans and the corresponding explanation features generated by LIME are listed in Table \ref{Tab-GrounTruth}. The results demonstrate that LIME can well capture the essential malicious behaviors of the malware samples. Since the malware samples in different malware families would conduct some similar behaviors such as sending messages, the SMS-related features are always generated as explanation features. However, for some families that contain unique malicious behaviors, they will contain some different explanation features. For example, \textit{fakeinst} samples usually install on other apps; thus, the INSTALL\_PACKAGES permission is their unique feature compared with the \textit{svpeng} and \textit{slembunk} samples.


\vspace{1em}
\noindent\fbox{
	\parbox{0.95\linewidth}{
		\textbf{Answer to RQ4}:
		For malware detection, the ranking of the five explanation approaches in term of the effectiveness metric is LIME $>$ LORE $>$ Anchor $\ge$ SHAP  $>$ LEMNA. For familial identification, the ranking is LIME $>$ SHAP $>$ Anchor $>$ LORE $>$ LEMNA. Furthermore, the identified effective explanation features can help us obtain the knowledge of most representative malicious behaviors for malware analysis.
	}
}

\subsection{RQ5: Can  the explanation approaches handle a great deal of samples?}
\label{subsec_exp_effciency}

\begin{table}[!t]
	\centering
	\scriptsize
	\caption{Runtime overhead (second) per sample for five explanation approaches on different datasets.}
	\scalebox{1.0}{
\begin{tabular}{cccccc}  \toprule[1.5pt]
	Dataset & LIME & Anchor & LORE  & SHAP    &  LEMNA\\ \hline
	MD-I    & 0.6  & 51.8   & 71.3  &  0.1    &  2.3\\
	MD-II   & 0.8  & 56.6   & 103.7 &  1.1    &  2.3\\
	MD-III  & 0.5  & 60.1   & 109.6 &  2.4    &  2.4\\
	MD-IV   & 0.8  & 126.7  & 167.1 &  4.5    &  2.1\\
	MD-V    & 1.1  & 63.4   & 85.2  &  1.7      &   2.6   \\
	FI-I    & 0.7  & 70.5   & 79.9  &  2.6    &  2.4\\
	FI-II   & 1.1  & 97.1   & 138.2 &  2.6    &  2.4\\
	FI-III  & 1.0  & 115.8  & 182.3 &  2.1    &  2.5\\
	FI-IV   & 1.0  & 156.1  & 216.4 &  2.6   &   2.6\\ 
	FI-V    & 1.6  & 140.9  & 177.3      &   2.8   &   2.6    \\ \bottomrule[1.5pt]
\end{tabular}
	}
\label{Tab-Runtime}
\vspace{-13pt}
\end{table}

\begin{figure*}[!t]
	\begin{minipage}[t]{0.32\linewidth}
		\centering
		\includegraphics[scale=0.32]{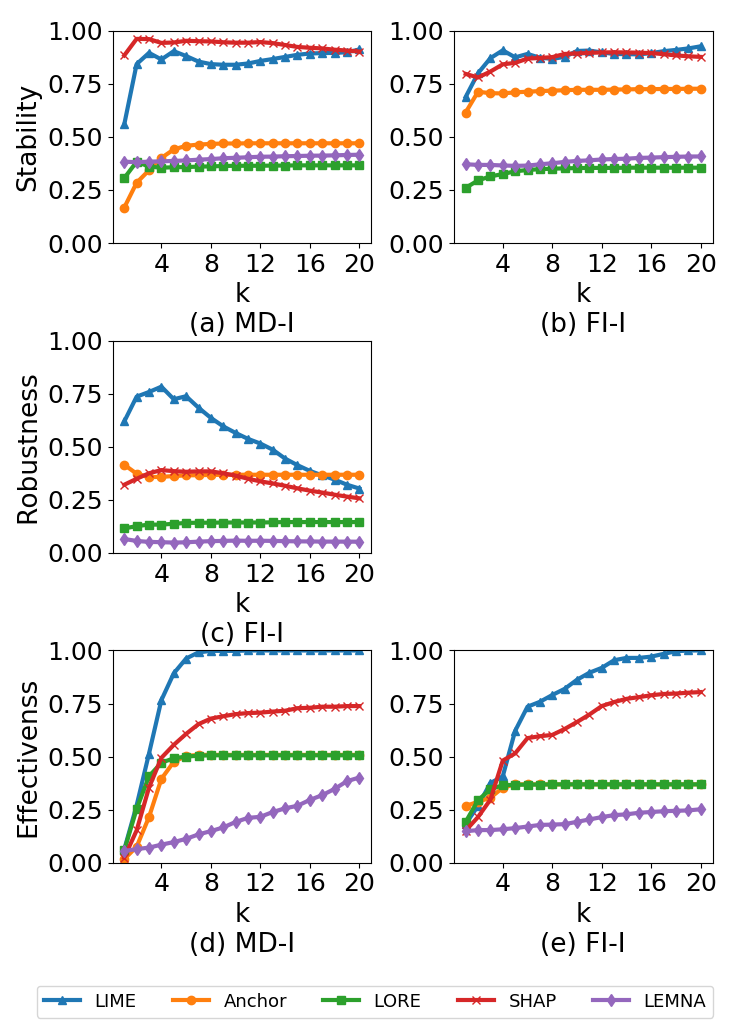}
		\caption{Evaluation results on SVM classifiers using three metrics.}
		\label{Fig-RQ6-SVM}
	\end{minipage}
	\hfill
	\begin{minipage}[t]{0.32\linewidth}
		\centering
		\includegraphics[scale=0.32]{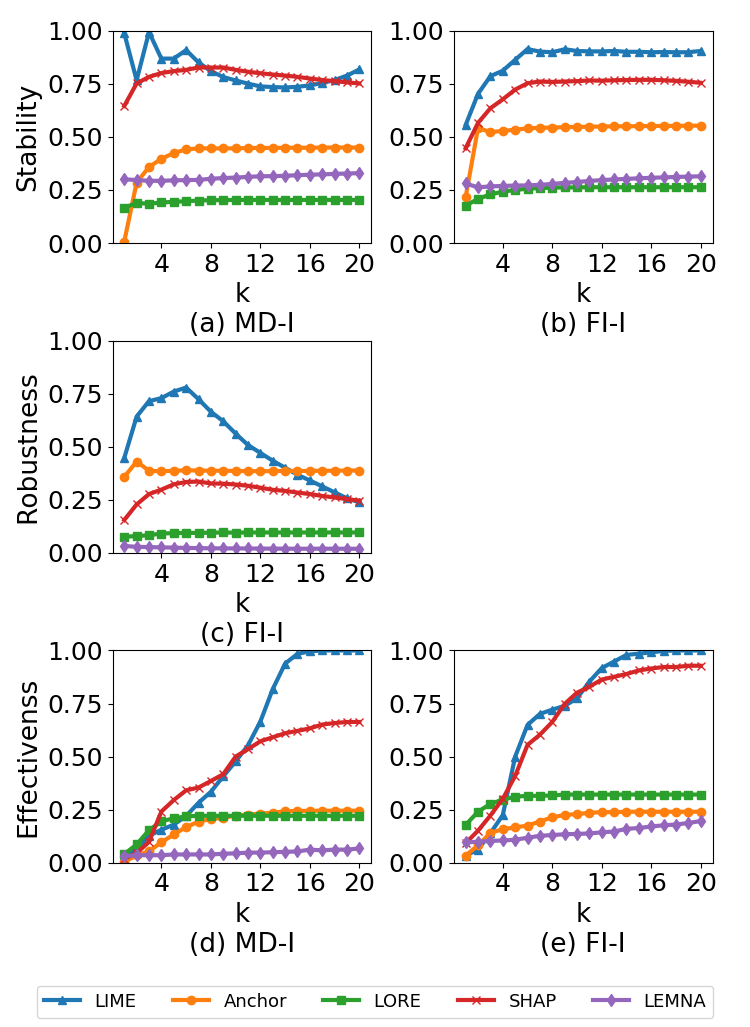}
		\caption{Evaluation results on KNN classifiers using three metrics.}
		\label{Fig-RQ6-KNN}
	\end{minipage}
	\hfill
	\begin{minipage}[t]{0.32\linewidth}
		\centering
		\includegraphics[scale=0.32]{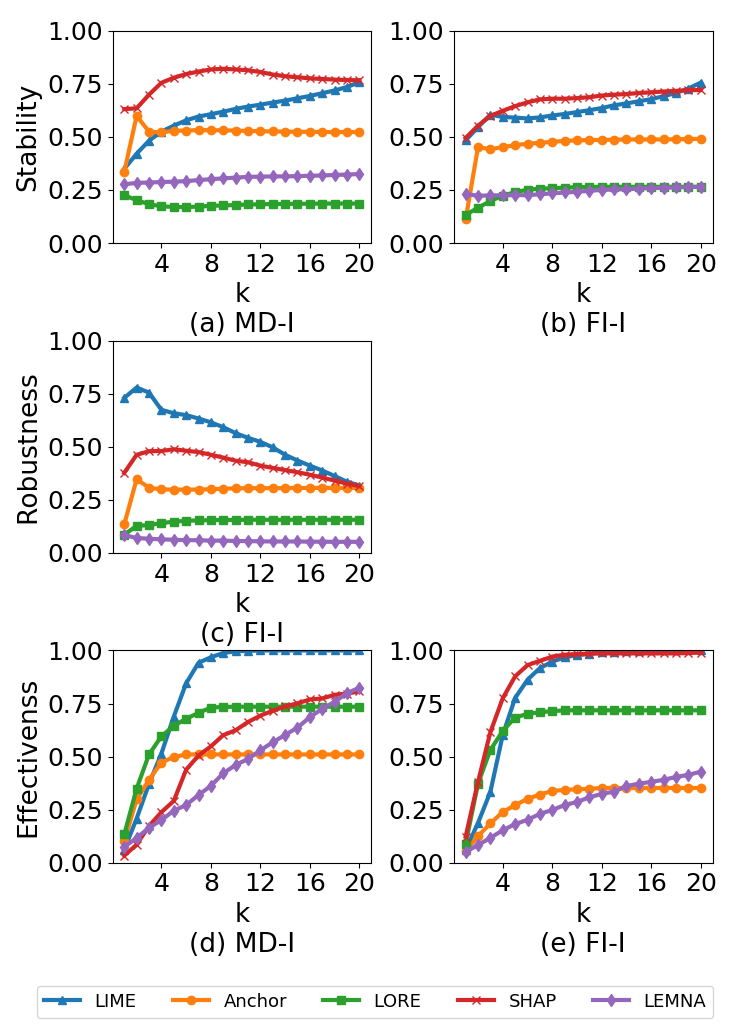}
		\caption{Evaluation results on MLP classifiers using three metrics.}
		\label{Fig-RQ6-MLP}
	\end{minipage}
\end{figure*}

To answer RQ5, we evaluate the runtime overhead for the explanation approaches on different datasets. The results are listed in Table \ref{Tab-Runtime}, where the number denotes the average seconds required to generate the explanation results for each sample.  Three observations are obtained from Table \ref{Tab-Runtime}.
\begin{itemize}
	\item {First, Anchor and LORE require much more time than LIME, SHAP, and LEMNA. For example, in FI-IV dataset, LORE and Anchor need 216.4 and 156.1 seconds on average to generate the interpretations while LIME, SHAP and LEMNA only need less than 3 seconds.}
	\item { Second, with the increase of the dataset size, the runtime overhead of Anchor and LORE increase. However, the runtime overhead of  LIME, SHAP, and LEMNA hardly change. }
	\item {Third, the runtime overhead of generating explanation results for familial identification is generally higher than that for malware detection.}
\end{itemize}

In summary, for malware analysis, LIME shows the best efficiency among the five  approaches. SHAP  and LEMNA present an acceptable efficiency. However, for Anchor and LORE, they are not suitable to be applied in malware analysis due to their high cost in extracting and reducing the interpretation rules.  Specifically, Anchor and LORE cost nearly two months to generate the explanation results for our datasets.

\vspace{1em}
 \noindent\fbox{
 	\parbox{0.95\linewidth}{
 		\textbf{Answer to RQ5}:
 		For malware analysis, the low efficiency of LIME, SHAP, and LEMNA make them be able to handle a large scale of malware samples while Anchor and LORE are not suitable in this domain.    
 	}
 }


%
%

\subsection{RQ6: Can  our metrics be evaluated on other classifiers?}
\label{subsec_exp_otherClassifier}

To answer RQ6, we evaluate the proposed metrics using three other classifiers (i.e., KNN, SVM, and MLP). Note that when evaluating the stability metric, we need to construct a set of similar models. Specifically, for KNN, we vary the number of neighbors from 8 to 11. For SVM, we vary the tolerance parameter as $1\mathrm{e}{-7}$, $2\mathrm{e}{-7}$, $3\mathrm{e}{-7}$ and $4\mathrm{e}{-7}$. For MLP, we vary the iteration numbers from 98 to 101.

The results are presented in Fig. \ref{Fig-RQ6-SVM}, Fig. \ref{Fig-RQ6-KNN}, and Fig. \ref{Fig-RQ6-MLP}. From the figures, we can observe that on different learning classifiers, the metric values for the same explanation approach are different. For example, the average stabilities of LIME and SHAP on SVM are much higher than those on MLP. However, even the metric values differ with classifiers; the comparison results among the five explanation approaches on the three metrics are nearly the same as those on the RF classifier. For stability, both LIME and SHAP can achieve good performance. For robustness, LIME can only perform well when k is less than 5. The other four approaches present weak robustness scores on all the classifiers. For effectiveness, LIME performs the best no matter what the classifier is. In summary, for malware analysis, LIME seems to be the most suited explanation approach. 

\vspace{1em}
\noindent\fbox{
	\parbox{0.95\linewidth}{
		\textbf{Answer to RQ6}:
		For malware analysis,our proposed three metrics can also be evaluated on other classifiers.    
	}

}

\section{Discussions and Threats to Validity}
\label{sec_validity}

\subsection{Discussions about the results}

In this section we discuss some insights about the evaluation results according to the designs of the five approaches. 

	For LIME, it builds a local linear model for an individual sample $x$ to approximate the boundary of prediction. To this end, it weights different perturbed samples according to the distances between $x$ and the perturbed samples. In this way,  LIME can, to some extent, restrict the domain potentially, from which the stability and the effectiveness of explanation results are relatively better~\cite{mittelstadt2019explaining}.  

Anchor provides explanations in the form of  a set of rules extracted incrementally from perturbations through a multi-armed bandit. However, the rules generated by Anchor are very
specific, and the average coverage is low.  It is likely to ignore the implicit impact of features that are less salient comparing with LIME, which may result in a loss of the explanation accuracy.

LORE obtains the explanation results based on the decision tree. However,  because of the inherent instability of the decision tree, LORE presents low performance.
Under the condition of the same model and parameters, the results of any two independent experiments are considerably different, indicating that two random 
perturbations produce almost different trees. Moreover, the correlation between features is ignored in the decision tree, which may be one of the reasons for its low performance.

For a certain feature, SHAP computes the shap values based on the average prediction differences of all possible feature combinations when its value flipped from 0 to 1.
This process satisfies three predefined desirable properties, local accuracy, missingness, and consistency that are grounded in game-theoretic principles. Therefore, 
SHAP presents relatively higher stability and effectiveness than other approaches.

LEMNA creates the mixture regression model to better approximate a non-linear decision boundary according to fused lasso. During the calculation, the model is presented in 
the form of a probability distribution, and parameters are estimated by using the EM (Expectation-Maximization) algorithm. However, the EM algorithm is sensitive to initial values, which is chosen randomly for each iteration. As a result, when we take a single instance for interpretation, consistency between explaining results is hard to guarantee, which might lead to the poor performance of LEMNA.  

Note that the robustness of all the above approaches is weak, which might be ignored by their designers. In recent years, some work can successfully attack explanation approaches.  The testing and enhancing of  explanation robustness would be a significant direction in future work. To mitigate the problem,  we think it is a promising way to combine the losing function of evaluation metrics when constructing the interpreters. For example, to construct an interpreter that has a high effectiveness score, designers can consider the label changing probability after feature mutation during the training phase.

\subsection{Threats to Internal Validity}

\noindent\textbf{Features provided by FeatureSmith.} We rely on the features provided by FeatureSmith in our work. The features are extracted from scientific papers. However, four years latter, the feature set might be outdated or incomplete, missing or incorrect features would make the generated interpretations biased.  For example, if the malware samples produced in recent years contain new malicious behaviors that are implemented with novel API calls, then the explanation approaches cannot capture such behaviors even they have good performance in old datasets. In future work, we plan to  reproduce the FeatureSmith in more published knowledge, not only the scientific papers published in recent years but also the technical blogs that are related with Android malware. 

\vspace{2mm}
\noindent\textbf{Limitation of Ground Truth Explanation Results.}
Different like the annotations generated by bounding box~\cite{szegedy2013deep} and semantic segmentation~\cite{long2015fully} in computer vision-related tasks, and the rational annotations~\cite{lei2016rationalizing} used in natural language processing, it is a time-consuming and error-prone job to manually construct ground truth annotations in Android malware analysis. In practice, even expert analysts cannot accurately identify all the malicious behaviors of malware samples. 
Although the authors of AMD dataset provide some descriptions about the malicious behaviors of each malware family, it is still a challenge to obtain the ground truth malicious behaviors for each sample, thus we cannot directly assess the quality of a single explanation result and compare the interpretability of the explanation approaches. However, our three general metrics measured on thousands of malware samples can assess the explanation approaches at a certain degree and give perceptual intuition to users which approach is more suitable.

\subsection{Threats to External Validity}

\noindent\textbf{Study on other explanation approaches.}
It is worth noting that our goal in this work is not to exhaustively evaluate all prior explanation approaches, but rather to highlight how our metrics apply to specific cases of interest, i.e., malware detection and familial identification.  Moreover, we believe that our proposed three metrics can also be used to conduct the sanity check for other explanation approaches since the metrics are the fundamental properties that an explanation approaches need to contain. The key point to measure other types of explanation approaches, e.g., model-specific approaches, is to select a proper scene that can construct the common classifier models. For example, the comparison between the CNN specific explanation approaches can be well conducted in the image classification scene. 

\vspace{2mm}
\noindent\textbf{Study on other domains.}
Malware analysis is an important problem that has close relations with our daily life and no previous work has been done for the measurement of explanation approaches in such critical domain.  We believe that our metrics can be generalized on other important domains such as sentiment analysis and image classification. The main difference between these domains is the feature representations. For malware analysis, the feature space is usually defined by analysts and they are represented as a numeric feature vector. However, for sentiment analysis, the features are generally the embedding vectors of words, and for image classification, the features are generally the pixel matrix. The key point to apply our metrics on other domains is replace the similarity calculation between the explanation features. For example, the explanation features of sentiment analysis are the sensitive words that have most impact to the decision-making. Thus, the similarity calculation should consider the words with close semantics. We think cosine distance between the embedding vectors obtained by \textsc{word2vec} is a promising way. It is worth noting that the ranking results concluded for malware analysis domain might be different from those in other domains.

\section{Related Work}
\label{sec_relatedwork}

\subsection{Android Malware Analysis}
Existing Android malware analysis approaches mainly rely on machine learning methods, of which the features can be categorized into two categories, string-based features and graph-based features. 
String-based features are mainly composed of request permissions, intents, API calls and  other components of the Android operation system. Arp \textit{et al.}~\cite{arp2014drebin} proposed Drebin, which performs a broad static analysis, gathering as many features of an app as possible such as permissions, API calls, and strings in Dalvik code. These features are then embedded in a joint
vector space for Android malware analysis. Aafer \textit{et al.}~\cite{aafer2013droidapiminer} proposed DroidAPIMiner, which extracts five different types of API calls as features. 
Graph-based features are proposed to improve the robustness of features while
retaining the program semantics. Zhang \textit{et al.}~\cite{zhang2014semantics} proposed DroidSIFT, which builds dependency graph databases and produces graph-based feature vectors by performing graph similarity queries.  Fan \textit{et al.}~\cite{fan2019graph} proposed GefDroid, which constructs SRA, a novel feature that depicts the similarity relationships between the structural roles of sensitive API call nodes in subgraphs, based on which they can perform a quick familial analysis. 

\subsection{Interpretable Explanation Approaches}
 In recent years, many explanation approaches have been proposed to provide interpretability for black-box models, using techniques based on local approximation, input perturbation, and back propagation. 
 
 Local approximation based explanation approaches are based on the
 assumption that the machine learning predictions around
 the neighborhood of a given input can be approximated by
 an interpretable white-box model. The five typical local explanation approaches are introduced before~\cite{ribeiro2016should, ribeiro2018anchors, guidotti2018local, lundberg2017unified,guo2018lemna}. 
 
 The input perturbation based approaches follow the philosophy that the contribution of a feature can be determined by measuring how prediction score changes when the feature is altered. Fong and Vedaldi~\cite{fong2017interpretable} proposed a meaningful perturbation approach, which learns  a saliency mask by blurring an image to minimize the probabilities of its target class. The saliency mask is the part of an image most responsible for a classifier decision. Li \textit{et al.}~\cite{li2016understanding} proposed a general methodology for
 interpreting neural network decisions by analyzing the effect of erasing particular representations. By analyzing the harm and the benefit the erasure does, the method provides a way to offer interpretable explanations and conduct 
error analysis on neural model decisions.  

Back propagation based approaches calculate the gradient~\cite{simonyan2013deep}, or its variants, of a particular output with respect to the input using back propagation to derive the contribution of features. Shrikumar \textit{et al.}~\cite{shrikumar2019not}  proposed DeepLIFT, which computes importance scores based on explaining the difference of the output from some ``reference'' output in terms of differences of the inputs from their ``reference'' inputs. Smilkov \textit{et al.}~\cite{smilkov2017smoothgrad} proposed SmoothGrad, which takes an image of interest, samples similar images by adding noise to the image, then takes the average of the resulting sensitivity maps for each sampled image as explanation results.

\subsection{Measurement of Explanation Approaches}
While many explanation approaches have been proposed,  however, several works have demonstrated that existing explanation approaches are easy to be attacked~\cite{ghorbani2019interpretation, heo2019fooling}. 
Slack \textit{et al.}~\cite{slack2019can} developed an approach that exploits the 
fact that LIME and SHAP are perturbation-based, to create a scaffolding around
any given biased black box classifier in such a way that its predictions on input data distribution remain biased, but its behavior on the perturbed data points is controlled to make the explanations look completely innocuous. Zhang \textit{et al.}~\cite{zhang2018interpretable} conducted a systematic study on the security of existing explanation methods and they proposed $ADV^2$, a general class of attacks that generate adversarial inputs not only misleading target DNNs but also deceiving their coupled interpretation models. 

Therefore, it is still an open question how to thoroughly validate the reliability of existing explanation approaches. In addition to the work~\cite{yeh2019fidelity,oana2019can,adebayo2018sanity} introduce before. 

The most related work is proposed by Warnecke \textit{et al.}~\cite{warnecke2019evaluating}, which is done without contacts in parallel. Both  of this work and ours focus on the evaluation of explanation approaches on security-related domains. Moreover, three common approaches are evaluated, i.e., LIME, SHAP, and LEMNA, and three similar measurement metrics are proposed, including stability, robustness, and efficiency. However, even both the two research groups have apparently worked on the same research question in parallel, there are three main differences. First, Even both \cite{warnecke2019evaluating} and our work focus on Android malware analysis domain. We evaluate on two tasks, i.e., malware detection and familial identification, while they only consider the former. Moreover, we conduct our evaluation on five different malware datasets which are constructed in different years, while paper~\cite{warnecke2019evaluating} uses two datasets. For the robustness metric, the paper~\cite{warnecke2019evaluating} assesses this metric based on the existing literature. They conclude that the explanation approaches are not robust just in a qualitative way, not in a quantitative way, as our definition. Third, we propose the effectiveness metric while \cite{warnecke2019evaluating} does not. The effectiveness is another fundamental property used to measure whether the explanation results are important to the decision-making.


\section{Conclusion}
\label{sec_conclusion}

In this paper, motivated by the inconsistency of explanation results generated by different explanation approaches, we evaluate their reliability problem in Android malware analysis.
We first propose principled guidelines to assess the quality of the explanation  approaches by designing three quantitative metrics to measure their stability, robustness, and effectiveness. Then we conduct a sanity check of five explanation  approaches in Android malware analysis based on our proposed metrics. The results demonstrate that our metrics can  not only assess the explanation approaches but also can help us obtain the knowledge of most representative malicious behaviors for malware analysis.



\bibliographystyle{IEEEtran}
\bibliography{IEEEabrv,mybibfile}

\end{document}